\documentclass[a4paper,fleqn]{cas-dc}

\usepackage[numbers]{natbib}
\usepackage[justification=centering]{caption}
\usepackage[caption=false]{subfig}
\usepackage{multirow}
\usepackage{wrapfig}
\usepackage{algorithm}
\usepackage[switch, pagewise]{lineno}
\usepackage[noend]{algpseudocode}

\begin{document}
\let\WriteBookmarks\relax
\def\floatpagepagefraction{1}
\def\textpagefraction{.001}
\shorttitle{SmartValidator: A Framework for Automatic Identification and Classification of Cyber Threat Data}
\shortauthors{Islam et al.}


\title [mode = title]{SmartValidator: A Framework for Automatic Identification and Classification of Cyber Threat Data}                      

\author[1,2]{Chadni Islam}
\ead{chadni.islam@adelaide.edu.au}
\author[1,2]{M. Ali Babar}
\ead{ali.babar@adelaide.edu.au} 
\author[1,2]{Roland Croft}
\ead{roland.croft@adelaide.edu.au} 
\author[2] {Helge Janicke}
\ead{helge.janicke@cybersecuritycrc.org.au} 

\address[1]{CREST – Centre for Research on Engineering Software Technologies, University of Adelaide, Australia}
\address[2]{Cyber Security Cooperative Research Centre, Australia}

\begin{abstract}
A wide variety of Cyber Threat Information (CTI) is used by Security Operation Centres (SOCs) to perform validation of security incidents and alerts. Security experts manually define different types of rules and scripts based on CTI to perform validation tasks. These rules and scripts need to be updated continuously due to evolving threats, changing SOCs' requirements and dynamic nature of CTI. The manual process of updating rules and scripts delays the response to attacks. To reduce the burden of human experts and accelerate response, we propose a novel Artificial Intelligence (AI) based framework, SmartValidator. SmartValidator leverages
Machine Learning (ML) techniques to enable automated validation of alerts. It consists of three layers to perform the tasks of data collection, model building and alert validation. It projects the validation task as a classification problem. Instead of building and saving models for all possible requirements, we propose to automatically construct
the validation models based on SOC's requirements and CTI.
We built a Proof of Concept (PoC) system with eight ML algorithms, two feature engineering techniques and 18 requirements to investigate the effectiveness and efficiency of SmartValidator. The evaluation results showed that when prediction models were built automatically for classifying cyber threat data, the F1-score of 75\% of the models were above 0.8, which indicates adequate performance of the PoC for use in a real-world organization. The results further showed that dynamic construction of prediction models required 99\% less models to be built than pre-building models for all possible requirements. Thus, SmartValidator is much more efficient to use when SOCs' requirements and threat behaviour are constantly evolving. The framework can be followed by various industries to accelerate and automate the validation of alerts and incidents based on their CTI and SOC's preferences.
\end{abstract}


\nonumnote{This preprint is accepted for publication in Journal of Network and Computer Applications, 2022.}

\maketitle

\section{Introduction}

Identifying and analyzing Cyber Threat Information (CTI) is an important part of validating security alerts and incidents~\cite{islam2019multi, koyama2015security, menges2019unifying, mittal2019cyber}. Any piece of information that helps organizations identify, assess, and monitor cyber threats is known as CTI~\cite{johnson2016guide}. To help a Security Operation Centre (SOC) in using CTI, existing approaches, such as a unifying threat intelligence platform~\cite{islam2019multi, koyama2015security, menges2019unifying, MISPJournal}, aim to automatically gather and unify CTI relevant to security alerts and incidents. However, gathering CTI is not enough to perform validation tasks, as security teams need to analyze and understand CTI for defining response actions. Security teams write scripts and define rules to extract necessary information from CTI, and map alerts and incidents to CTI~\cite{anstee2017great, elmellas2016knowledge, RFID2021, tounsi2018survey}. Whilst techniques such as defining rules and scripts can be automated, they do not help in identifying evolving threats and alerts ~\cite{serketzis2019actionable, ward2017building, zhou2019ensemble}, because rules can only be defined for behavior of known threats. Thus, human understanding and resolution are required to identify, define and update CTI, rules and scripts for emerging threats to adapt changing contexts.

The vast volume of CTI makes it time-consuming for a human to analyze. Thus, to address the shortcoming of defining rules and scripts to use CTI, we present a novel framework, SmartValidator. SmartValidator identifies CTI and validates security alerts and incidents by leveraging Artificial Intelligence (AI) based automation techniques~\cite{faiella2019enriching,  qamar2017data,serketzis2019actionable}. SmartValidator follows a systematic and structured approach for reducing the human cognitive burden to continuously monitor for changes (e.g., change in attack patterns and CTI) and define the automation strategies whenever changes occur. We focus on two aspects of automation: (i)~\textit{automatic identification} of CTI for different alerts and (ii)~\textit{automatic validation} of alerts using identified CTI. 

By \textit{automatic identification} of CTI, we mean identifying CTI from a wide variety of sources. The increasing presence and amount of CTI over the internet demands effective techniques to automate the identification of the required CTI for validation tasks~\cite{faiella2019enriching,menges2019unifying,noor2019machine,qamar2017data}. The sources of CTI vary with differences in alerts and incidents~\cite{EY2017,MISP2021,RFID2021}. Examples of CTI include Indicators of Compromise (IoC) (system artifacts or observables associated with an attack), Tactics Techniques Procedures (TTP) and threat intelligence reports~\cite{johnson2016guide}. 

By \textit{automatic validation} of alerts and incidents, we refer to validating (i.e., prioritizing, and assessing the relevance or impact of) different types of alerts and incidents generated and identified by different detectors. In this context, by detector we mean any tools or systems used for detection of malicious activities. An organization deploys or develops different types of detectors that generate alerts upon detection of malicious activities. Examples of such detectors include Intrusion Detection Systems (IDS), vulnerability scanners and spam detectors. Validation of different types of security alerts and incidents requires extracting information from relevant CTI~\cite{anstee2017great, elmellas2016knowledge, RFID2021, tounsi2018survey, WINKLER2017143}. For example, a network administrator or threat hunter writes scripts to search for CTI (e.g., information about the suspicious incident) and defines rules to validate an alert. 
There are always cases for which automatic validation would not be suitable. For example, in our scenario, automated validation is not applicable for alerts and incidents that do not have associated CTIs; hence, such scenario would require a security team to perform manual analysis.

The massive volume and variations of CTI opens the door for automatic identification of patterns and gathering insights about CTI using Natural Language Processing (NLP) and Machine Learning (ML) techniques. For instance, Sonicwall has reported 9.9 billion malware attacks in its 2020 cyber threat report~\cite{sonicwall2020}. The threat research team of Sonicwall has come across more than~1,200 new malware variants each day. Existing studies~\cite{ibrahim2020challenges, le2019automated, noor2019machine,RFteam2018,  serketzis2019actionable, Struve2017, zahedi2018empirical, zhou2019ensemble} have highlighted  the power of AI to monitor, gather and analyze security intelligence. Recent advances have also been noticed in the use of NLP and ML techniques to extract patterns from threat data and gain insight about attacks and threats. The focus of these studies are application-specific, for example, detecting anomalies~\cite{zhou2019ensemble} or automating vulnerability assessment~\cite{le2019automated}, which need to be updated with changing CTI and organizational needs. These studies required knowledge of NLP and ML to build a model for performing the assessment or detection task. Most existing SOCs are managed SOCs (SOC as a service), which are subscription based~\cite{Nick2020, ibrahim2020challenges}. They do not have dedicated data science or ML experts to design and update the AI based system based on their need. Considering this scenario \textit{“Can we design an efficient system to automate and assist the validation of security incidents and alerts with changing threat data and user needs”?}.

Evolving threat landscapes and changing needs of security teams demand a dynamic AI/ML-based validation system which can be adapted at runtime.
For instance, if a security expert expresses interest to validate the maliciousness of a domain "URL", a prediction model is built by a data scientist team that classifies a URL as malicious or non-malicious. In an ML context, this task is known as a prediction or classification task. We propose three different layers to differentiate the tasks of threat data collection, validation and prediction model building. The purpose is to hide implementation complexity of data processing, prediction models and validators from security teams. Each layer is controlled and developed by experts with dedicated capabilities. Changing threat landscapes require SOCs to request new CTI and prediction models. One possible solution to this is to build and save prediction models for all possible attributes sets. However, building all possible models whenever changes occur will incur significant resource consumption (e.g., computation time). Most SOCs have limited resources; hence, instead of pre-building all possible combinations of prediction models, SmartValidator is designed to build the model based on a SOC's demands.

We have implemented a Proof of Concept (PoC) system to evaluate the effectiveness and efficiency of our proposed framework with two feature engineering approaches and eight ML algorithms. We have used an IoC form of CTI~\cite{johnson2016guide} and collected open source threat intelligence (OSINT) from public websites and a CTI platform, MISP~\cite{MISP2021, RFID2021}. The input of the developed system is a set of attributes from a security team. For example, a security team may want to investigate the “\textit{domain name}” and “\textit{URL}” to identify the “\textit{maliciousness}” of an incident. The developed system takes these three attributes as input where “\textit{domain name}” and “\textit{URL}” are the observed attribute~($ob_{attrib}$) and “\textit{maliciousness}” is the unknown attribute~($un_{attrib}$). The prediction models are built for classifying/ predicting~$un_{attrib}$ based on~$ob_{attrib}$. 

To capture changing contexts (i.e., security team requirements), we have considered five $un_{attrib}$: (i)~attack, (ii)~threat type, (iii)~name, (iv)~threat level and (v)~event. Eighteen different sets of~$ob_{attrib}$ (shown in Table~\ref{tab:tableAttribute}) are provided to validate these five attributes to demonstrate the performance of the PoC with changing requirements. We have designed the PoC to select the suitable feature engineering approaches and ML algorithms at run time. 
Seven $ob_{attrib}$ are selected by the PoC to predict \textit{attack} and 11 $ob_{attrib}$ sets are used to predict the remaining four attributes. Hence, the PoC provided a total 51 optimal prediction models for predicting five $un_{attrib}$ based on the preferred~18~$ob_{attrib}$. The results show that approximately~84\% of the models have F1-scores above~0.72 and~75\% of the models have F1-scores above~0.8. These results imply that SmartValidator is able to assist the automatic validation of threat alerts with a high level of confidence. Most of the models that were built with data gathered from the MISP platform can effectively predict~$un_{attrib}$ based on~$ob_{attrib}$ with a higher F1-score than the models that were built with CTI gathered from public websites. This demonstrates that trusted threat intelligence is more effective in validating alerts.

The results also demonstrate the efficiency of SmartValidator with dynamic changes in the preferred set of attributes. We pre-built all possible models, which required us to run~814 experiments. Given a maximum time limit of 48 hours and a memory limit of~100GB to build each prediction model, 20\%~of the models failed to complete within the time limit and given memory. Hence, it shows the difficulties a security team would encounter in manually constructing each model. Results further reveal that building prediction models is a time-consuming process and requires expertise that can be automated through orchestrating different tasks. Saving the feature engineering approaches and ML algorithms helps SmartValidator to use them for predicting new attributes based on changing CTI and SOC requirements. Thus, constructing prediction models at run time based on a security team's preferred attributes sets reduces the overhead and resource consumption. The key contributions of this work are:

\vspace{-5pt}

\begin{itemize}
\item A novel AI-based framework, SmartValidator, that consists of three layers to effectively and efficiently identify and classify CTI for validating security alerts with changing CTI and security team requirements.
\vspace{-5pt}
\item A PoC system that automatically built 51 models to predict five different unknown attributes with~18~ observed attributes sets using two sources of OSINT.
\vspace{-5pt}
\item We demonstrated that SmartValidator can effectively select optimal prediction models to classify CTI where approximately 75\% of optimal models have an F1-score of above~0.8.
\vspace{-5pt}
\item We showed the efficiency of SmartValidator by building prediction models based on security team demands which requires approximately 99\%~less models to build, thus less resources and time consumption. 
\vspace{-5pt}
\end{itemize}


Paper organization: Section~\ref{Sec:Motiv} presents a motivation scenario that highlights the need for SmartValidator. Section~\ref{Sec:Preli} discusses the background knowledge about CTI. Section~\ref{Sec:Framework} introduces the proposed framework, SmartValidator. Section~\ref{section:POC} describes the large-scale experiment that is carried out for the evaluation of SmartValidator. Section~\ref{Sec:Eval} demonstrates the effectiveness and efficiency of the proposed approach. Section~\ref{Sec:RelatedWork}  discusses related work. Finally, section~\ref{Sec:Conclu} concludes the paper with future works.
\vspace{-10pt}

\section{Motivation Scenario} \label{Sec:Motiv}

\begin{figure*}
    \centering
        \subfloat[]{\includegraphics[width=0.7\textwidth]{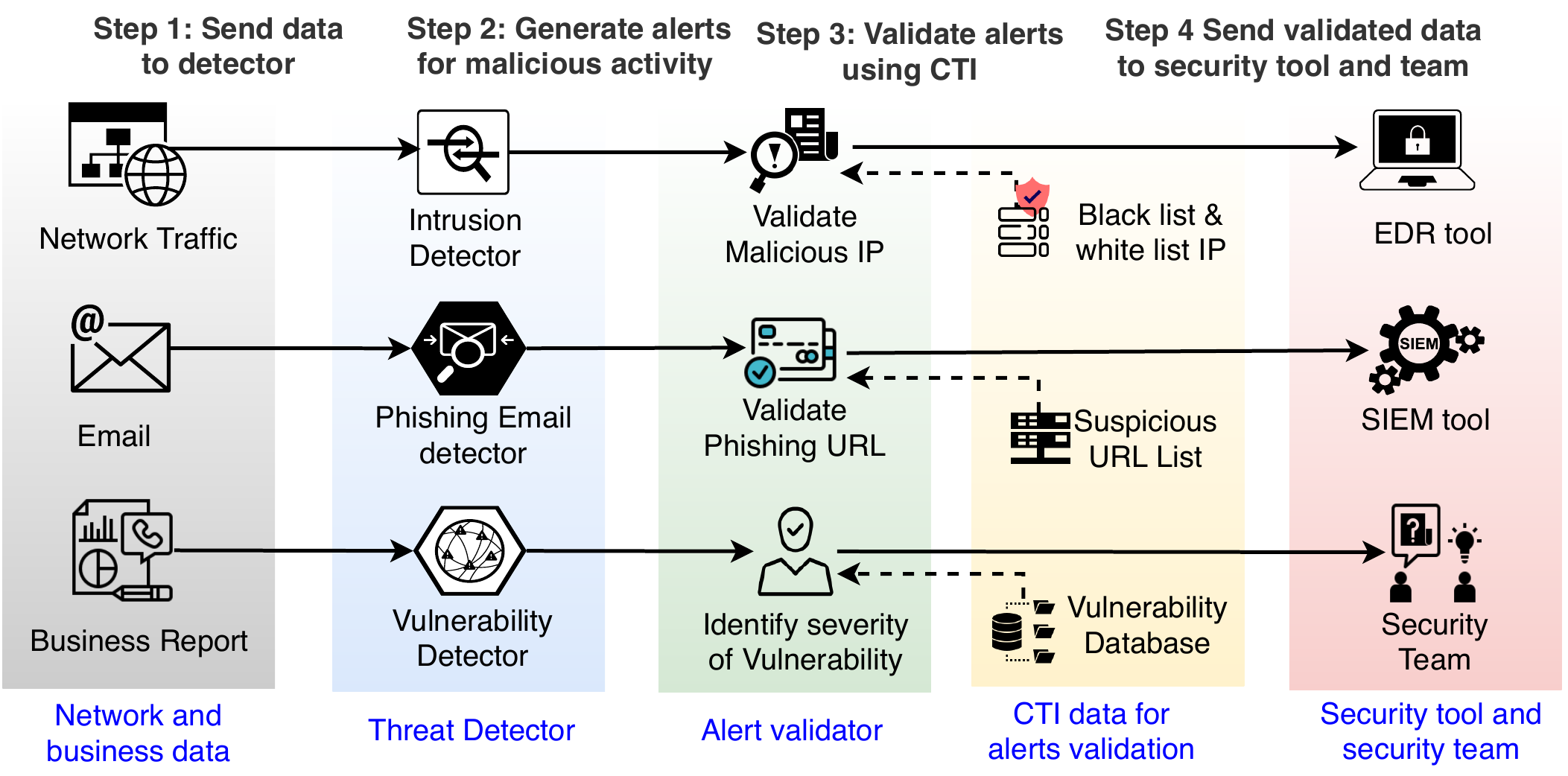}\label{fig:motiva}}
        \hfil
        \subfloat[]{\includegraphics[width=0.8\textwidth]{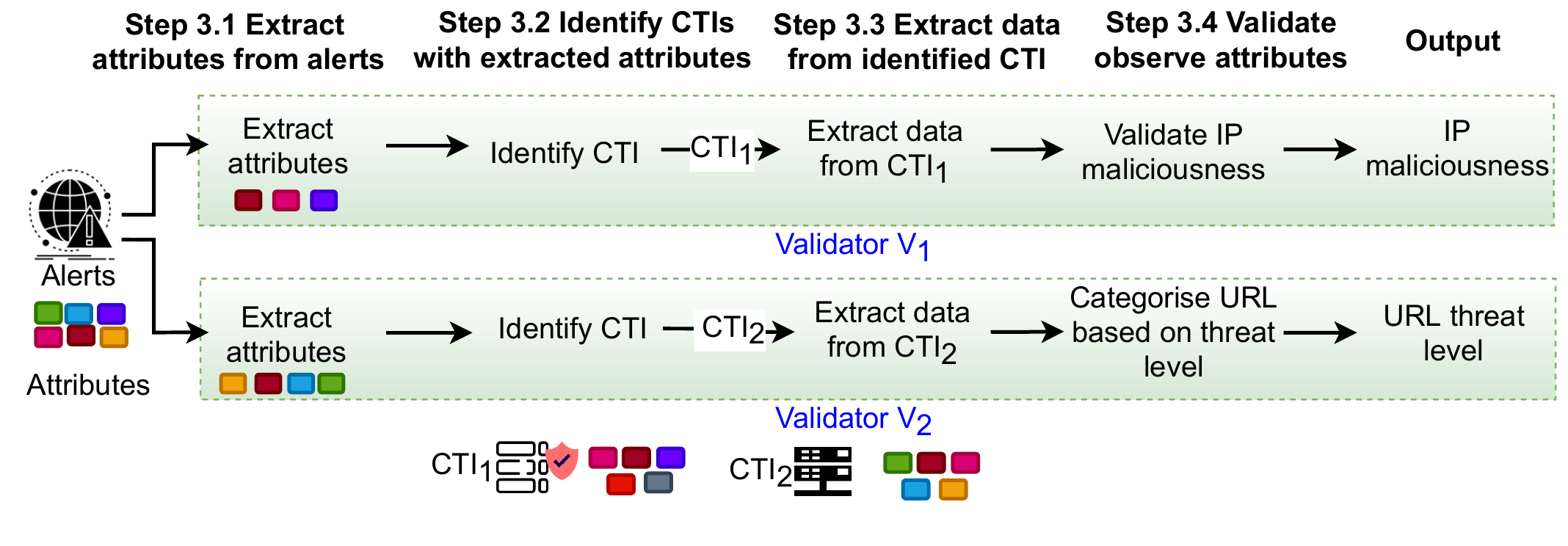}\label{fig:motivb}}
    \caption{Motivation scenario illustrated (a) detection of malicious activities and validation of alerts with multiple detectors and sources of CTI respectively, (b) validation of same alerts for different set of preferences required two different validators and different CTI}
    \label{fig:Motiv}
    \vspace{-15pt}
\end{figure*}

In this section, we motivate an AI-based solution for alert validation through an example scenario. Figure~\ref{fig:Motiv} shows a scenario where a SOC of an organization has deployed different types of detectors, validators and CTI to monitor and validate malicious behaviour in its network and business data. Figure~\ref{fig:motiva} shows that three detectors (intrusion, phishing email, and vulnerability detectors) are deployed to detect suspicious and malicious activities of an organization. The information used by detectors varies with attack types. For example, the information that detectors use to identify an intrusion is different from identifying a phishing email\footnote{https://github.com/counteractive/incident-response-plan-template/blob/master/playbooks/playbook-phishing.md} (Figure~\ref{fig:motiva}). These detectors continuously monitor an organization's network and business data\footnote{https://github.com/rosenbet/demisto/tree/master/Playbooks} (e.g., emails, network traffic and business reports). 

Most detectors produce alerts upon detecting malicious activity that require a security team to act on it. These alerts require validation before analysing them for decision making\footnote{https://www.incidentresponse.com/playbooks/}. In this paper, we consider a validator performs a task related to prioritizing and identifying the relevance or impact of alerts. Let us assume that an intrusion detector has detected a list of malicious IP addresses. A SOC has an alert validator to validate the maliciousness of the IPs~\cite{Siemplify2019}. Figure~\ref{fig:motiva} shows that validation of different types of alerts requires different forms of CTI. To validate IP maliciousness, blacklist and whitelist IP addresses are used. CTI further varies from organization to organization. Each security team has its own set of requirements (different attributes) to validate an alert. In this scenario, the security team rely on three types of CTI for alert validation. Considering different types of alerts have different attributes and require different sources of CTI, a SOC needs three validators to validate the alerts produced by three detectors. Figure \ref{fig:motivb} illustrates the use of CTI to validate alerts.

We assume, an alert of type $A_i \in A$ ($A$ is the set of alerts) is produced by detector $D_1$ (e.g., IDS). Each alert is represented with different attributes (or features). The function $F_{attrib}$($A_i$) provides attributes list of $A_i$. 

$F_{attrib}(A_i) \rightarrow <f_1, \ f_2,\ f_3,\ f_4,\ f_5>$
\\ where  $f_1 = \text{IP}$, $f_2 = domain$, $f_3= \text{URL}$, $f_4 = attack\ type$, and $f_5  = threat\ level$. Considering two different security roles of a SOC have different preferences and used different attributes to validate $A_i$, two validators $V_1$ and $V_2$  are built (Figure \ref{fig:motivb}). For validator $V_1$, a security team prefers to validate \textit{attack type} based on \textit{\text{IP}}, \textit{domain} and \textit{\text{URL}}, thus

$ob^1_{attrib} = <\text{IP},\ domain,\ \text{URL}>$ and 

$un^1_{attrib} = <\text{IP}\ maliciousness>$. \\
 For validator $V_2$, a security team's preference is to validate \textit{threat level} using attributes \textit{domain}, \textit{URL} and \textit{attack type}, thus,

$ob^2_{attrib}= <\text{IP},\ domain,\ \text{URL},\ attack\ type>$ and 

$un^2_{attrib}= <\text{URL}\ threat\ level>$.

For both cases, to perform validation, validators first extract $ob_{attrib}$  and if available $un_{attrib}$ from $A_i$ and then identify CTI with these attributes. In most cases, a security team provides CTI sources to a validator. In the next step, CTI that have $ob_{attrib}$ and $un_{attrib}$ are identified. As shown in Figure \ref{fig:motivb}, validator $V_1$ extracts three attributes to validate \textit{IP maliciousness} and validator $V_2$ extracts four attributes to validate \textit{URL threat level}. $\text{CTI}_1$ has the attributes that are required by validator ${V_1}$. On the other hand, $\text{CTI}_2$ has the attributes required to investigate URL threat level by $V_2$. Therefore, threat data is extracted from $\text{CTI}_1$ and $\text{CTI}_2$ respectively for further investigation.

Though $V_1$ and $V_2$ have investigated two different sets of attributes, the key steps (step~3.1 to step~3.4), as shown in Figure~\ref{fig:motivb}, are the same. The tasks $V_1$ and $V_2$ can be formulated as an ML classification problem, where two different prediction models are required to be built. Building a prediction model involves pre-processing of data (e.g., $ob^1_{attrib}$), feature engineering, training and selecting a model, and then predicting a output ($un^1_{attrib}$). Many of the possible $ob_{attrib}$ of Cyber Threat Information (CTI) (e.g., domain, filename, description and comment) are textual features. Traditional categorical feature engineering or transformation approaches are not suitable to encode the textural features, hence require the application of NLP technique. To perform validation of $un^1_{attrib}$ and $un^2_{attrib}$ using $ob^1_{attrib}$ and $ob^2_{attrib}$, prediction models are required to be built where input of prediction models are security team preferences. Here, we consider the observed attributes and unknown attributes as SOCs' preferences/ requirements. We assume, $AS$ as a set of a SOC's requirements, where 

 $AS =  <ob_{attrib},\ un_{attrib}> $
\\ This mean for validator $V_1,\ AS_1$ =  <$ob^1_{attrib}$ , $un^1_{attrib}$> \& $AS_1 \in AS$. For $ V_2, AS_2$ =  <$ob^2_{attrib}$, $un^2_{attrib}$> \& $AS_2 \in AS$

Considering emerging threat patterns, a SOC may deploy new detectors and update the existing rules of intrusion detectors to detect evolving anomalies. To validate the alerts of a new detector, new validators may be required. Thus, several changes can arise in the scenario of Figure~\ref{fig:Motiv}. Following, we present the three scenarios that we consider in this work.

\vspace{-5pt}

\begin{itemize}
    \item Change in Alert: With changes in alerts types, variation can be seen in the attributes of alerts. For example, an alert of type $A_2$ may have a different set of attributes from $A_1$ such as timestamps, date, IP, organization, tools and comments. Depending on types of attack and detector, an alert attribute changes. Building prediction models with changing alerts might require incorporation of various types of pre-processing and features engineering approaches. 
    \vspace{-5pt}
    \item Change in CTI: CTI are continuously changing with changing requirements from SOCs. In most cases, SOCs buy CTI from third parties where attributes provided by different vendors vary, or they build their own CTI platform. The validation of alerts relies on the attributes available in CTI.
    \vspace{-5pt}
    \item Changes in Preferred Attributes Set: Change in preferred attributes ($AS$) requires re-designing and building of prediction models. A SOC does not always have dedicated data scientists or experts to design and build prediction models. Even though the steps of model building are repetitive (e.g., pre-processing, feature engineering, model building and selection), few changes may be required for adaptation of the variations as existing solutions are not designed to automatically work with changing attributes.
\end{itemize}
\vspace{-5pt}
To address these changes, we propose SmartValidator to support the flexible design of a validator following a systematic and structured approach. The proposed framework can automatically construct prediction models to validate alerts
with changing requirements. 

\vspace{-10pt}

\section{Preliminaries} \label{Sec:Preli}
This section provides background information about CTI and MISP (an Open Source Threat Intelligence Platform).

\textbf{Indicator of Compromise: } IOCs provide characteristics of cyberattacks and threats. Based on IOCs, a security team decides whether a system is affected by a particular malware or not~\cite{ anstee2017great, elmellas2016knowledge, tounsi2018survey}. Examples of IOCs include domain name, IP, file names and md5 file hashes. Three common categories of IOCs are network indicators, host-based indicators and email indicators. IP addresses, URLs and domain names are the most popular types of network indicators. The malicious file hash or signature, registry keys, malware name and dynamic link libraries are widely used host-based indicators. An email indicator may consist of a source email address, message objects, attachments, links and source IP addresses. The source of IOCs ranges from crowd-sourcing to government-sponsored sources. Just having threat data is not enough to fully understand the context or patterns of a cyberattack. For example, threat data may contain an IP address that is used only once to attack a network. Conversely, an associated URL in threat data might have been used many times. Therefore, threat intelligence must be extracted from the threat data with possible IOCs and their contextual information~\cite{ anstee2017great, elmellas2016knowledge, ward2017building, tounsi2018survey}. Table~\ref{tab:CTIsource} shows examples of some of these websites that provide threat feeds and are utilized for gathering OSINT. Table~\ref{tab:CTIexample} shows examples of CTI publicly available in the malware domain website~\cite{malwareDomain2021}. 

\begin{table*}[h]
 \caption{Description of each CTI sources with the Indicator of Compromise (IOCs) they contain}
    \begin{tabular}{ll}
  \toprule
    \textbf{Source} & \textbf{Description}\\
   \midrule
C\&C Tracker \cite{Ctracker2019} & Contains a list of C\&C IPs (command  and control botnets), date, and a link to a manual which contains \\&  text description and false positive risk value. \\
Feodo Tracker \cite{Ftracker2019} & Tracks the Feodo Trojan. Contains IP, port, and date. \\
Malware Domain List \cite{malwareDomain2021} &  Contains a searchable list of malicious domains, IP, date, domain, reverse lookups \\ & and lists registrants. Mostly focused on phishing, Trojans, and exploit kits.\\
Ransomware Tracker \cite{Ransomwware2019} & Provides overview of infrastructures used by Ransomware, status of URLs, IP address and \\ & domain names associated with Ransomware and various block list of malicious traffic. \\
WHOIS data \cite{WhoIS2019} & Provides a database of registered users and assignees of internet resources, which is widely used\\ &  for lookup of domain names.\\
Zeus Tracker \cite{Ztracker2019} & Tracks domains of Zeus Command \& Control servers.\\
OpenPhish \cite{OpenPhish2019} & A list of phishing URLs and their targeted brand. \\
   \bottomrule
    \end{tabular}
    \label{tab:CTIsource}
\vspace{-10pt}
\end{table*}

\begin{table*}[!hb]
\caption{An example of a list of observed malware domains with corresponding Indicator of Compromise (IOCs) from the website malwaredomainlist.com \cite{malwareDomain2021}}
    \centering
    \begin{tabular}{llllll}
    \toprule
       \textbf{Date} & \textbf{Domain} & \textbf{IP} & \textbf{Reserve Lookup} & \textbf{Description} & \textbf{ASN}\\
\midrule      
       
2017/12/04 18:50 & textspeier.de    & 104.27.163.228 & - & phishing/ fraud & 13335 \\
2017/10/26 13:48 & photoscape.ch/ & 31.148.219.11 & knigazdorovya.com  & trojan & 14576 \\
       &Setup.exe \\
2017/06/02 08:38 & sarahdaniella.com/swift & 63.247.140.224 & coriandertest. & trojan & 19271\\
  & /SWIFT\%20\$.pdf.ace & & hmdnsgroup.com & \\
  
2017/05/01 16:22 & amazon-sicherheit.kunden & 63.247.140.224 & hosted-by.blazingfast.io & phishing  & 49349\\
& -ueberpruefung.xyz \\
2017/03/20 10:13 & alegroup.info/ntnrrhst & 185.61.138.74 &  mccfortwayne.org & Ransom, Fake & 197695\\
&&&&.PCN, Malspam \\
   \bottomrule
   \end{tabular}
    \label{tab:CTIexample}
    \vspace{-10pt}
\end{table*}

\textbf{Threat Intelligence}: Threat Intelligence, also known as security threat intelligence, is an essential piece of CTI for a cybersecurity team. According to \textit{Recorded Future}, “\textit{threat intelligence is the output of the analysis based on detection, identification, classification, collection and enrichment of relevant data and information}”~\cite{RFID2021}.Threat intelligence helps a security team understand what causes an attack and what needs to be done to defend against it by gathering contextual information about an attack. For example, security teams use threat intelligence to validate security incidents or alerts and enrich threat data to get more insights about a particular security incident~\cite{anstee2017great, elmellas2016knowledge, tounsi2018survey, RFID2021, WINKLER2017143}. The gathered data is organized in a human-readable and structured form for further analysis~\cite{anstee2017great, elmellas2016knowledge, faiella2019enriching, RF2019,  WINKLER2017143}. Open Source Intelligence (OSINT) is gathered from various websites (e.g., Zeus Tracker and Ransomware Tracker) that provide information about malware or blacklist domain/IPs. 




\textbf{Cyber Threat Intelligence Platform:} Threat intelligence platforms allow security communities to share and collaborate to learn more about the existing malware or threats. Using threat intelligence platforms, companies can improve their countermeasures against cyber-attacks and prepare detection and prevention mechanisms. In recent years, the cybersecurity communities have emphasized building common threat intelligence platforms to share threat information in a unified and structured way, and make CTI actionable~\cite{faiella2019enriching, gao2018graph, menges2019unifying,mittal2019cyber, tounsi2018survey, ward2017building}. Various specifications and protocols such as STIX, TAXII, Cybox, CWE, CAPEC and CVE are widely used to describe and share threat information  through common platforms~\cite{ Stixbarnum2012standardizing, barnum2012cybox, taxiiconnolly2014trusted, RamsdaleCTISurvey2019}. Trusted Automated Exchange of Indicator Information (TAXII)~\cite{taxiiconnolly2014trusted} is developed as a protocol for exchanging threat intelligence represented in STIX format. Both STIX and TAXII are open source and have collaborative forums~\cite{Stixbarnum2012standardizing,taxiiconnolly2014trusted}.  

\begin{figure}
     \includegraphics[scale = 0.35]{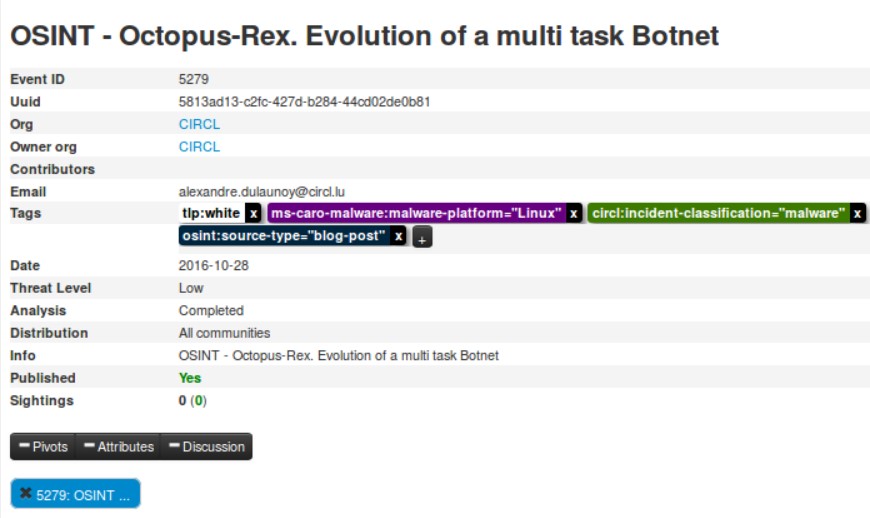}
    \caption{An example of MISP showing evoluation of a multi task Botnet}
    \label{fig:MISP}
    \vspace{-15pt}
\end{figure}

\textbf{MISP:} Malware Information Sharing Platform (MISP) is one of the most popular trusted threat intelligence platforms used by different industries to store and share CTI~\cite{MISP2021, MISPJournal}. MISP is a framework for sharing and storing threat data in a structured way~\cite{MISPJournal, azevedo2019pure}. MISP enables an organization to store both technical and non-technical information about attacks, threats and malware in a structured way. The relationship between malware and their IOCs are also available in MISP. Rules for network Intrusion Detection Systems (IDS) can also be generated from MISP, which can be imported into an IDS system, and hence improve the detection of anomalies and malicious behavior. A security team queries MISP for relevant data, and it shows the details of the attack. For example, Figure~\ref{fig:MISP} shows the details of the evolution of a multitask Botnet. Table~\ref{tab:MISPExample} and Table~\ref{tab:MISPvaluePercentage} of Appendix~\ref{app:A} show the key attributes gathered from MISP for this study and and the percentage of each attribute.

\vspace{-10pt}

\section{Proposed Framework} \label{Sec:Framework}
\begin{figure*}
    \centering
     \includegraphics[width=0.85\textwidth]{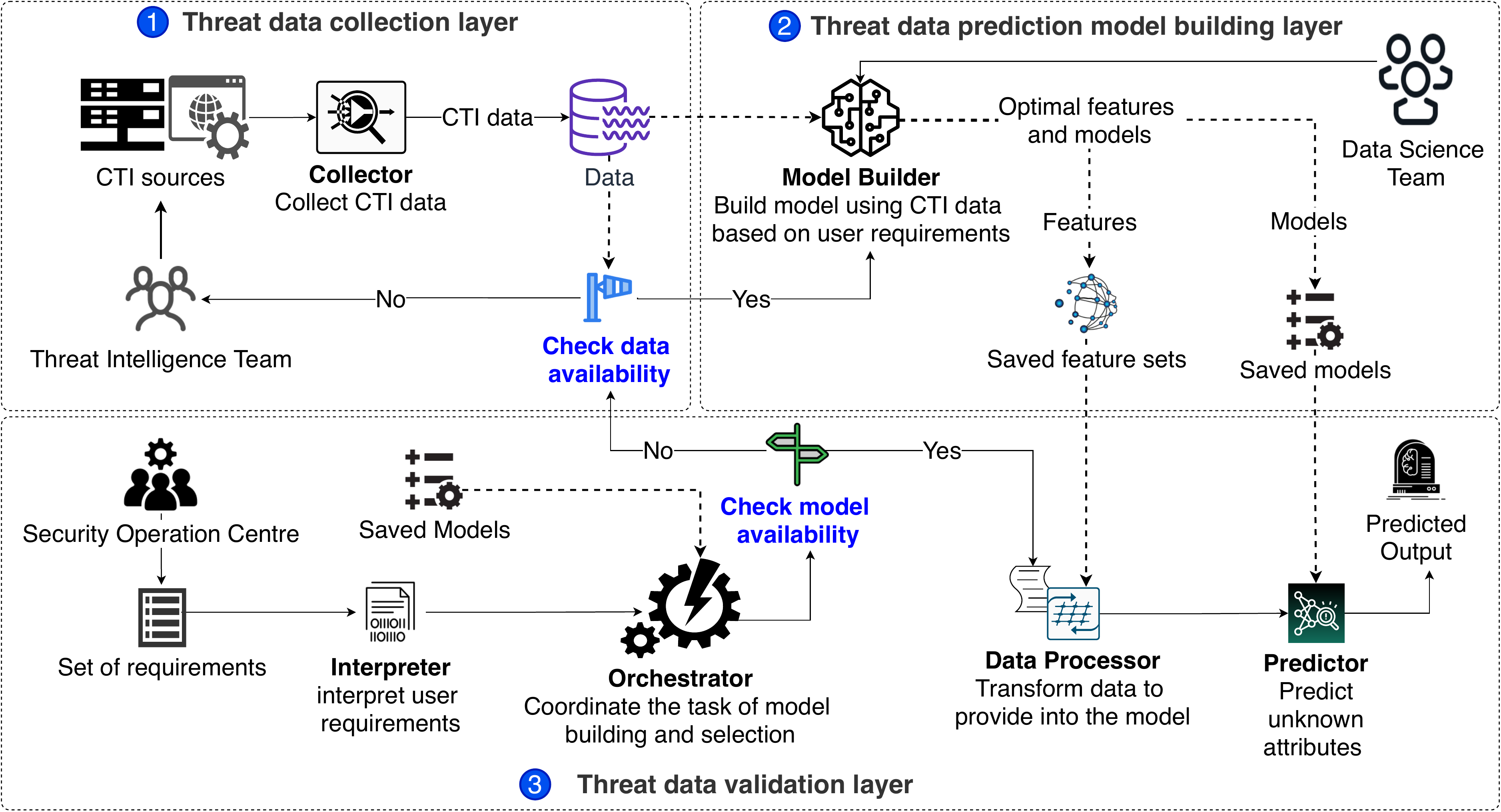}
    \caption{An overview of SmartValidator for automated identification and validation (i.e., classification) of cyber threat data}
    \label{fig:framework}
    \vspace{-15pt}
\end{figure*}

Figure~\ref{fig:framework} provides an overview of our proposed framework, SmartValidator, that automates the identification and classification of CTI for validation of alerts. It comprises of three layers: (i)~threat data collection layer, (ii)~threat data prediction model building layer and (iii)~threat data validation layer. We consider each layer to have a separation of concerns so that while updating components of one layer, a security team does not need to worry about other layers.

\textit{Threat Data Collection layer} (section~\ref{sub:4.1}): The threat data collection layer consists of a collector that automates the identification of CTI information from various sources (Figure~\ref{fig:datalayer}). It further transforms the gathered CTI into a unified form (if required) and passes it to the threat data prediction model building layer.

\textit{Threat Data Prediction Model Building layer}(section~\ref{sub:4.2}): This layer has a model builder that builds models for validation of alerts based on a SOC's requirements ($ob_{attrib}$, $un_{attrib}$) using the gathered CTI (Figure~\ref{fig:predictionlayer}). Pre-processing of data, feature engineering, training and evaluation of prediction models are performed in this layer to generate candidate validation models. The candidate models and corresponding feature sets are saved with a SOC's preferences for use by the threat data validation layer at runtime to validate alert. 

\textit{Threat Data Validation layer} (section~\ref{sub:4.3}): The threat data validation layer takes alerts and a SOC's requirements as input to choose suitable prediction models for alert validation. SOC requirements are interpreted by an interpreter. Based on a SOC's requirements, attributes are extracted from alerts. The extracted attributes are used to choose suitable prediction model from the candidate list of saved models for predicting the unknown attributes ($un_{attrib}$) and perform validation of alerts based on observed attributes ($ob_{attrib}$).

The following sections elaborate the core components and functionalities of each layer of SmartValidator. 
\vspace{-5pt}

\subsection{Threat Data Collection Layer}\label{sub:4.1}
\begin{figure}[htb]
     \includegraphics[width=0.5\textwidth]{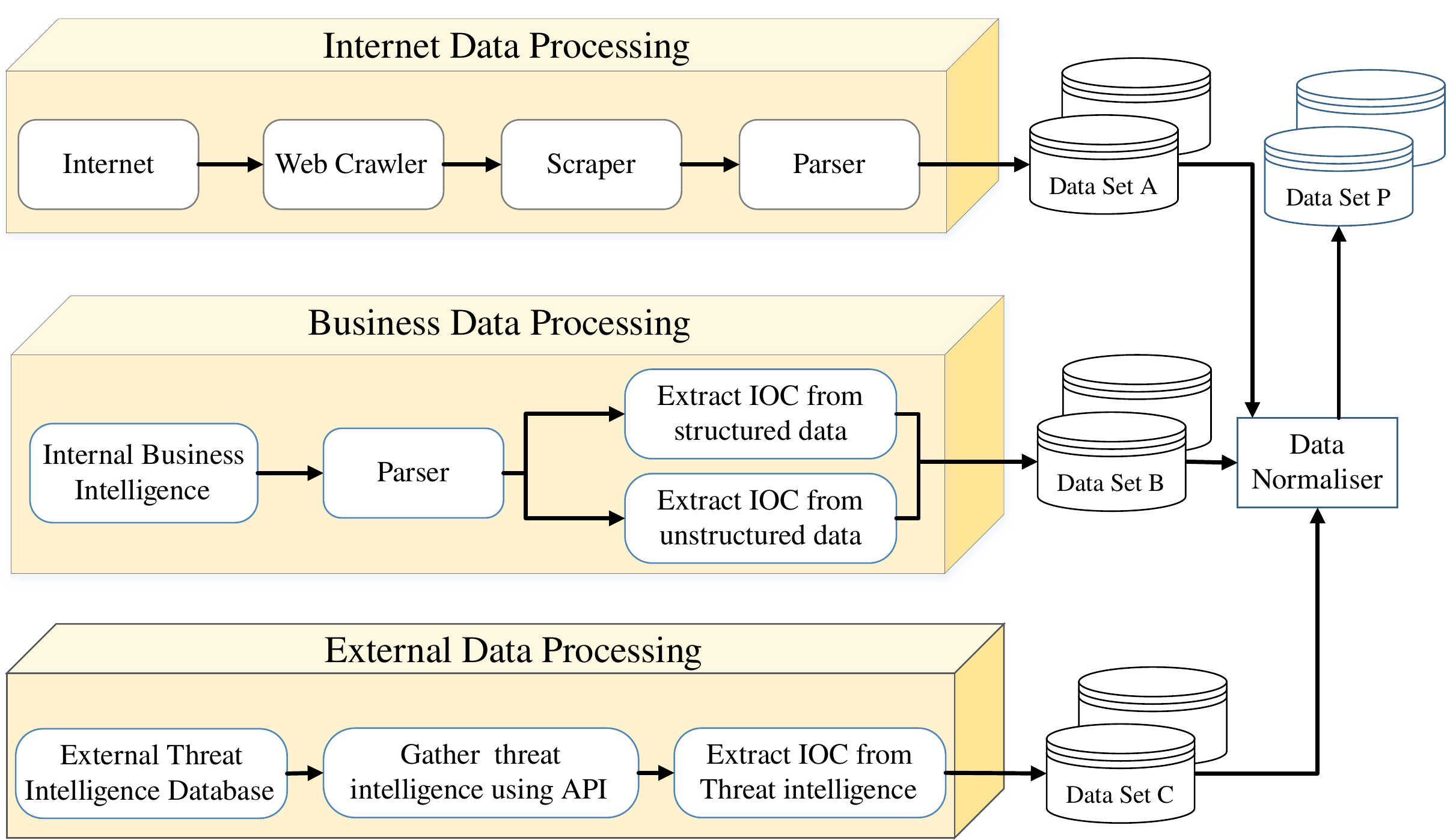}
    \caption{Threat data collection layer - IOCs are extracted and processed from three types of data and then combined into a unified form}
    \label{fig:datalayer}
    \vspace{-15pt}
\end{figure}

Validation of a security alert requires the identification of relevant CTI for building a threat data prediction model. The purpose of the prediction model is to learn the pattern of CTI for automatic validation of alerts. Here, we have formulated the validation tasks as a classification task. For instance, to validate an IP maliciousness, a system needs to classify IPs as malicious or non malicious, which can be achieved through a prediction model. Similar to existing studies~\cite{faiella2019enriching, tounsi2018survey, MISPJournal}, the data collection layer gathers CTI from multiple sources and combines them into a unified format. The data collection layer employs a collector to gather CTI data from an organization’s preferred sources. Considering several types of CTI sources are used by an organization, deployment of various plugins, APIs and crawlers are required to collect CTI from these sources. Figure~\ref{fig:datalayer} has shown the processing of three types of CTI data -  (i)~internet data, (ii)~business data and (iii)~external data. These are most commonly used CTI. Other forms of CTI can also be integrated by following standard data processing strategies. 

For processing internet data, the collector has web crawlers, scrapers and parsers to gather and process CTI data from web pages (Figure~\ref{fig:datalayer}). A web crawler searches and identifies reliable sites that contain threat information and IOCs of various malware. Considering threat intelligence team  provides the relevant list of websites or keywords of their interest to gather CTI~\cite{listThreatIntel2021}, a crawler crawls through the internet to search the relevant information. We propose a scraper as part of the collector for filtering out unnecessary information from crawled data. Crawling and scraping can be done on a variety of sources, such as RSS feeds, blogs and social media, but require different types of processing. A parser utilizes various information processing techniques to extract information from the output of a scraper and organise the data into a structured and language-agnostic format (e.g., markup language such as ontology). For forums or blog posts written in \textbf{natural language}, a parser is required to extract threat information from sentences. NLP tools and techniques (e.g., Spacy and NLTK) are used to build a parser based on the structure of a document and information required by security team. 

To get threat feeds from databases and CTI platforms, we design API calls and queries that are parts of the collector. Threat feeds can be gathered from both an organization’s internal business data and external data (Figure~\ref{fig:datalayer}). In this paper, we only gather external threat feeds. The collector can also query external data sources to find out missing information about available threat data. For example, after receiving an IP address, a query can be made to WHOIS query website to search for domain name. In this way, a collector gathers different sets of data, for example, blacklist and white list IP addresses, list of phishing websites and so on, from different types sources. The collected data is further combined into a unified form (e.g., dataset~P, Figure~\ref{fig:datalayer}). To combine the data into a unified form, we first normalised the data, removed redundant information and then combined them. Examples of normalisation techniques include 1NF, 2NF and so forth. Depending on validation tasks (e.g., validation of IP maliciousness or validation of domain threat level), CTI is extracted and sent to the threat data prediction model building layer to build a validation model.

We consider organizations (e.g., government or financial) may have a dedicated threat intelligence teams or may use third party services to gather CTI. Any update related to the collection of CTI, such as adding or modifying CTI sources, deploying APIs or parsers to gather and extract information from these CTI, and inclusion or deletion of new data  collection and normalisation techniques, are performed in the threat data collection layer by dedicated threat intelligence team.
\vspace{-5pt}

\subsection{Threat Data Prediction Model Building Layer}\label{sub:4.2}
The threat data prediction model building layer is designed to build ML-based classification models using CTI and SOC's preferences. For example, if a security team wants to validate the \textit{maliciousness} of an IP considering the \textit{IP}, \textit{domain} and \textit{URL}, an ML classification model is built that takes IP, domain and URL and predicts IP maliciousness. We consider these attributes (IP, domain, URL and IP maliciousness) as SOC's preference where $ob_{attrib} = \{\textit{IP}, domain, \textit{URL}\}$ and $un_{attrib} = \{\textit{IP}\ maliciouness\}$. SOC's preferences drive from organizational security requirements and alerts. 

\begin{figure} [tbh]
     \includegraphics[scale = 0.5]{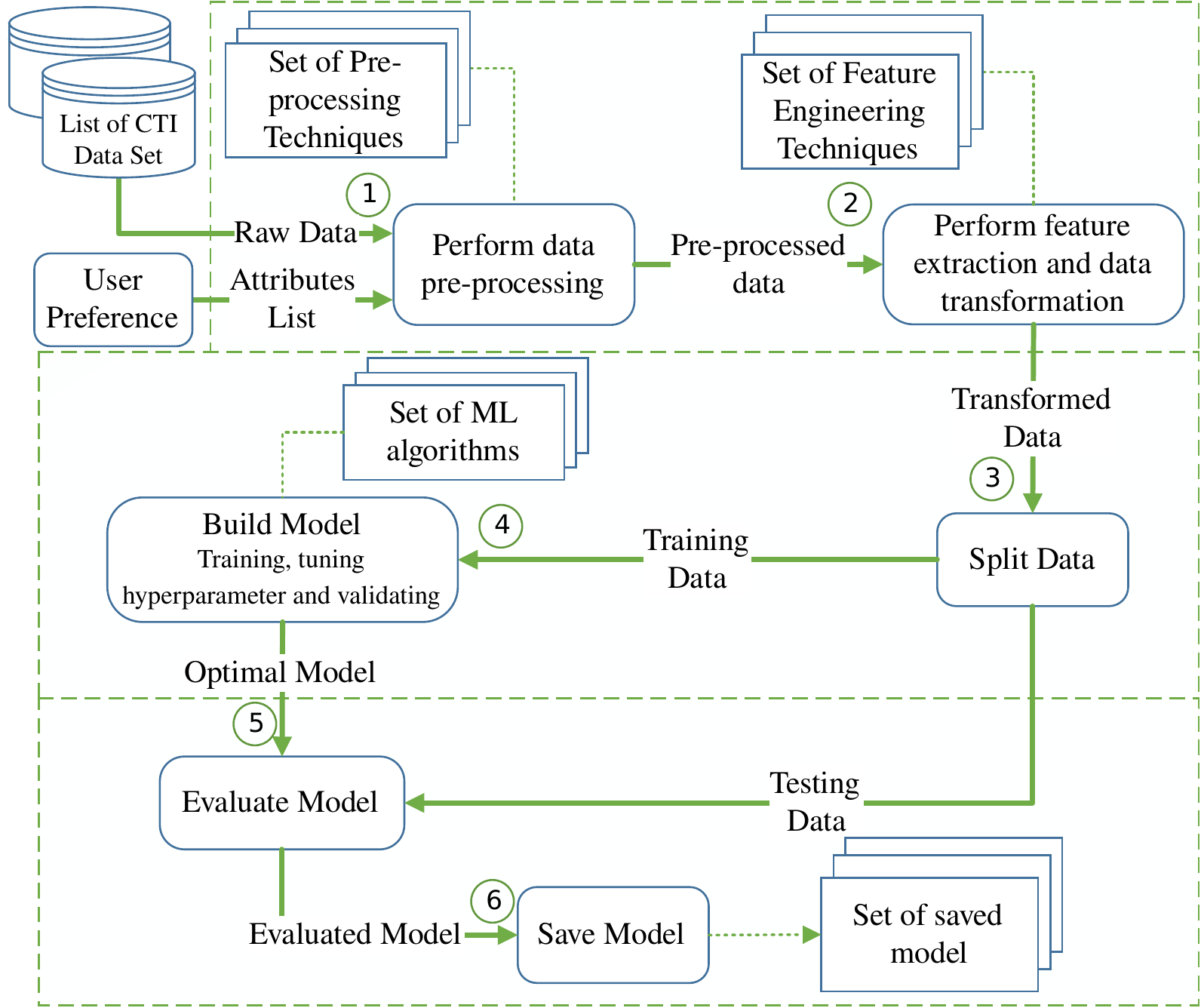}
    \caption{Threat data prediction layer - prediction models are built to predict unknown attributes based on observed attributes}
    \label{fig:predictionlayer}
    \vspace{-15pt}
\end{figure}

Figure~\ref{fig:predictionlayer} shows the core components and workflow of the threat data prediction model building layer. It comprises of a pre-processor that pre-processes CTI (step~1, Figure~\ref{fig:predictionlayer}) for extracting features from it. The pre-processing techniques depend on the types of attributes (e.g., categorical or text-based). For example, IP maliciousness can be either malicious or non-malicious, which is categorical. On the other hand, domain name is text based, as shown in Table~\ref{tab:CTIexample}.  Pre-processed data is passed to a feature engineering module where data is transformed into features (numeric form) (step 2, Figure~\ref{fig:predictionlayer}), which are used as input for the ML algorithms.  The reason behind this is that ML algorithms can only work with numerical data~\cite{Helge2020, RFteam2018, sabir2020machine, Struve2017}. Depending on the type, size and diversity of CTI, the data science team chooses a feature engineering approach. The first two steps of Figure~\ref{fig:predictionlayer}~leverage simple NLP techniques for pre-processing and feature engineering. Categorical values can be directly transformed into features using label encoding or one-hot encoding and text values are transformed into features using count vectorizer and TFIDF (Term Frequency-Inverse Document Frequency) techniques. The associated text cleaning and pre-processing steps for each text attribute are discussed in Appendix B1. Common pre-processing techniques such as tokenization, stop words removal and lemmatization are performed before transforming text data into features. Hence, based on the alert attributes types, data science teams perform data pre-processing and select a feature engineering approach. 

The transformed data is split into training and testing datasets to build, select and evaluate an prediction model (step~3, Figure~\ref{fig:predictionlayer}). ML algorithms are applied to train models based on CTI that connects and learns patterns in the data to derive a working model. Depending on the nature of the training data, different models are built by a data science team. Traditionally a set of ML algorithms are applied to find an algorithm suitable for a specific dataset and user requirements~\cite{AHMED201619, Helge2020, sabir2020machine}. To investigate the effectiveness of prediction models for the validation task, we considered a set of ML algorithms (e.g., Decision Tree, Naïve Bayes, K-Nearest Neighbours and Random Forest). The details of the PoC is discussed in section~\ref{section:POC}. As most ML algorithms have a list of hyperparameters, validation techniques (e.g., k-fold cross-validation, random cross-validation and Bayesian optimisation) are incorporated to select  hyperparameters\footnote{Hyperparameters are user-defined values that determine details about the ML classifier before training. For example, a decision tree requires tuning the value of variable depth, and k-nearest neighbours has a variable number of neighbours.} setting and feature engineering approach for a specific ML algorithm (step~4, Figure~\ref{fig:predictionlayer}). 

The built model’s performance is evaluated using the testing dataset (step~5, Figure~\ref{fig:predictionlayer}). Different types of performance metrics (e.g., precision, recall, accuracy and F1-score) are used to choose a model (also known as the optimal model) that provides the best performance (details in section~\ref{sub:evaluationMetrics}). In this work, we mainly consider F1-score for evaluation, which is a score between~0 and~1. Higher F1-score indicates better performance of a model. Figure~\ref{fig:predictionlayer} shows how the pre-processing techniques, feature engineering approaches and ML algorithms that are used to build threat data prediction models are stored for future model building process. Once an prediction model is found to have performed best, that model can be rebuilt using both training and testing datasets. The best models are saved with the evaluation score (step~6, Figure~\ref{fig:predictionlayer}) for using it in the threat data validation layer.

\subsection{Threat Data Validation Layer}\label{sub:4.3}
We design the threat data validation layer to (i)~collect a SOC's needs, (ii)~automatically orchestrate and request for CTI and prediction models and (iii)~validate alerts.
Figure~\ref{fig:framework} shows the validation layer comprises of an  \textit{interpreter}, \textit{orchestrator}, \textit{data processor} and \textit{predictor}. In this layer, security teams of SOCs provide their preferences, $AS$, as a set of requirements, where $AS$ = {<$ob_{attrib}, un_{attrib}>$}. Security teams may also provide a minimum threshold for F1-score, which we refer to as the confidence score of prediction models. The reason behind gathering confidence score is that the performance of prediction models will differ with variation in CTI, alerts and attributes sets. A security team might need a higher F1-score while dealing with safety critical data and sensitive information. For example, to identify IP maliciousness, a security team may request for a validation model with a minimum confidence score of 0.9. While categorizing comments or text messages as spam, model performance of 0.8 or above may be approved. Selection of F1-score values vary from application to application. Thus, instead of setting a fixed value, we consider providing security teams the flexibility to set the confidence score based on their application needs.  

We design Algorithm~\ref{alg:Orchestrator} describing the key steps of the threat data validation layer. These steps are coordinated and orchestrated by the orchestrator. SOC preferences ($AS$ and confidence score) are the input of Algorithm~\ref{alg:Orchestrator}. The \textit{interpreter} receives the SOC requirements and extracts observed attributes $ob_{attrib}$, unknown attributes $un_{attrib}$ and confidence scores from that (line~3). The orchestrator checks model availability with F1-score above the confidence score for predicting $un_{attrib}$ based on $ob_{attrib}$ (line~4). If a model is available, the attributes are passed to the data processor, where it pre-processes and transforms the data based on the saved pre-processing and feature engineering approaches (lines~5-7). Finaly, the pre-processed and transformed data is sent to the predictor, which uses the available model to predict $un_{attrib}$ (line~8).

\begin{algorithm}[tbh]
\footnotesize
    \caption{Model building with orchestrator in threat data validation layer}\label{alg:Orchestrator}
    \begin{algorithmic}[1]
        \State \textbf{Input: } \texttt{AS}  <$ob_{atrrib}, un_{attrib}$>, \texttt{confidence score}
        \State \textbf{Output: } \texttt{predictedData}
        \State Interpret (\texttt{AS, confidence score})
        \State  \texttt{IsModels}= CheckModel(\texttt{AS, confidence score}) 
        \If{\texttt{IsModels} true}
            \State \texttt{model, featureEng} = getModel(\texttt{AS,confidence score})
            \State \texttt{processedData} = transformData(\texttt{featureEng}, \texttt{AS})
            \State \texttt{predictedData} =  predictOutput(\texttt{model,processedData})
         \Else
            \State IsData = CheckData(\texttt{AS})
            \If{IsData true}
                \State \texttt{CTIData} = RetrieveData(\texttt{CTI, AS})
                \State  \texttt{model} = buildModel(\texttt{CTIData, AS, confidence score})
                \If{\texttt{model} is built}
                    \State go to step 6
                \Else
                    \State go to step 19
                \EndIf
            \Else
                \State RequestData(\texttt{AS})
                \State \textbf{return NotApplicable} 
            \EndIf
        \EndIf
        \State \textbf{return} \texttt{predictedData}
     \end{algorithmic}
\end{algorithm}

If a model is unavailable for the requested attributes set, e.g., for predicting~$un_{attrib}$ based on~$ob_{attrib}$, then the orchestrator requests the data collector module to gather the relevant CTI data for the preferred attribute sets (line~9). After identifying the relevant information, the data collector module sends the collected CTI data to the \textit{model builder} to build an prediction model (lines~12-13). Using the CTI \textit{data} from the data collection layer, the \textit{model builder} follows the model building process as discussed in section~\ref{sub:4.2} (lines~14-15). After building a model, it sends the model availability notification to the orchestrator. Then the same process of data processing and prediction is performed. If the requested data is unavailable, a notification is sent to a threat intelligence team and a SOC team to gather the required CTI and manually analyze the alerts, respectively (lines 18-19). To ensure that alerts are not ignored when models or CTIs are unavailable, SOC teams must keep informed that manual analysis is required.

Our proposed framework, SmartValidator, streamlines gathering, identification and classification of CTI. SmartValidator allows a security team to swiftly make a response about incoming alerts. As most information is generated in a structured way, it can be easily pre-processed to share through a CTI platform such as MISP or Collective Intelligence Framework (CIF) to benefit diverse security teams. SmartValidator can be integrated with the existing security orchestration and automation process to validate alerts and thus work together with the existing security tools, such as Security Information and Event Management (SIEM) and Endpoint Detection and Response (EDR). Microsoft Azure Sentinel\footnote{https://azure.microsoft.com/en-au/services/azure-sentinel/} and Splunk\footnote{https://www.splunk.com/} are examples of SIEM where Limacharlie\footnote{https://www.limacharlie.io/} and Google Chronicle\footnote{https://cloud.google.com/blog/products/identity-security/introducing-chronicle-detect-from-google-cloud} are considered as EDR.

\vspace{-10pt}
\section{Experiment Design and Setup} \label{section:POC}

We designed and implemented  a Proof of Concept (PoC) system to evaluate SmartValidator. We expected to demonstrate effectiveness of prediction models in validating security threat data and efficiency of building prediction models based on a SOC's requirements. The goal of the PoC is to identify the relevant CTI and build prediction models based on a list of SOC's requirements. Hence, we evaluated the PoC system based on the following two Research Questions (RQ).  
\vspace{-5pt}
\begin{itemize}
    \item RQ1. How effective is machine learning in classifying CTI for SmartValidator?
   \vspace{-5pt}
    \item RQ2. How efficient is SmartValidator in selecting and building prediction models at runtime over pre-building all possible prediction models?
\end{itemize}
\vspace{-5pt}
To build the PoC system for SmartValidator, we implemented the core components of Figure \ref{fig:framework}: the data collection layer presented in Figure \ref{fig:datalayer}, the prediction layer presented in Figure \ref{fig:predictionlayer} and an orchestrator for the validation layer described in section \ref{sub:4.3}. 

\subsection{SOC's Requirement}

We defined a set of attributes (validated attributes and observed attributes) as SOC's requirements to carry on the experiment. These attributes were mainly given by a team who were different than the one that implemented the prediction models. Thus, here we considered the team who provided the requirement as part of the SOC and the other team as part of the data science team. This setting gave us the option to evaluate a variety of different SOC's requirements, to appropriately assess the PoC system. In a practical scenario, these requirements would usually be defined by a security team. Among the various attributes, commonly validated attributes ($un_{attrib}$) are \textbf{attack}, \textbf{threat type} and \textbf{threat level}. We considered them as the desired unknown attributes. Beside them, we also considered two others attributes \textbf{name} and \textbf{event} as the desired unknown attributes.
Example of these attributes are shown in Table~\ref{tab:MISPExample} in appendix~\ref{app:A}. As shown in Table \ref{tab:MISPExample}, an example of an event title (or event) is “\textit{OSINT Leviathan: Espionage actor spear phishes maritime and defence targets}”.

As observed attributes ($ob_{attrib}$) vary from security team to security team, we gathered~18 different sets of observed attributes from security team to validate the aforementioned $un_{attrib}$. As shown in Table~\ref{tab:tableAttribute}, \textit{IP information} (i.e., ASN, IP owner, country, and domain), \textit{organization}, \textit{comments about attributes}, \textit{comments about attacks}, \textit{event data}, \textit{timestamp} and \textit{category} are the attributes that we considered in the set of observed attributes. We selected these attributes from alert data that are also commonly used to validate alerts generated by different IDS. We also used the metadata of attributes such as URL, domain and filename. 
\begin{table}[!hb]
\caption{List of the observed attribute sets}
    \centering
    \begin{tabular}{ll}
    \toprule
       \textbf{\#} & \textbf{List of attributes}\\
\midrule      
   $ob^1_{attrib}$ & Date \\
   $ob^2_{attrib}$ & Domain\\
   $ob^3_{attrib}$ & IP, ASN, Owner, Country \\
   $ob^4_{attrib}$ & Date, Domain \\
   $ob^5_{attrib}$ & IP, ASN, Owner, Country, Domain \\
   $ob^6_{attrib}$ & IP, ASN, Owner, Country, Date \\
   $ob^7_{attrib}$ & IP, ASN, Owner, Country, Domain, Date \\
   $ob^8_{attrib}$ & IP destination, Port, IP source, ASN, Owner, \\ & Country, Domain, File hash, Filename\\
   $ob^9_{attrib}$ & IP destination, Port, IP source, ASN, Owner, \\ & Country, Domain, Description, Comment, File \\ & hash, Filename\\
   $ob^{10}_{attrib}$ & IP destination, Port, IP source, ASN, Owner, \\ & Country, Domain, Description, Comment\\
   $ob^{11}_{attrib}$ & IP destination, Port, IP source, ASN, Owner,\\ & Country, Domain, Date, Timestamp, File hash , \\ & Filename\\
   $ob^{12}_{attrib}$ & IP destination, Port, IP source, ASN, Owner, \\ & Country, Domain, Date, Timestamp, Description,\\ &  Comment, File hash, Filename\\
   $ob^{13}_{attrib}$ & IP destination, Port, IP source, ASN, Owner, \\ & Country, Domain, Date, Timestamp, Description,\\ &  Comment\\
   $ob^{14}_{attrib}$ & IP destination, Port, IP source, ASN, Owner, \\ &Country, Domain, Date, Timestamp\\
   $ob^{15}_{attrib}$ & Description, Comment, File hash, Filename\\
   $ob^{16}_{attrib}$ & Date, Timestamp, File hash, Filename\\
   $ob^{17}_{attrib}$ & Date, Timestamp, Description, Comment, \\ &File hash, filename\\
   $ob^{18}_{attrib}$ & Date, Timestamp, Description, Comment\\
   \bottomrule
   \end{tabular}
    \label{tab:tableAttribute}
    \vspace{-10pt}
\end{table}

\subsection{Collecting CTI}\label{subsec:5.2}

We gathered CTI from two types of sources – publicly available internet data and data from an OSINT platform, MISP. CTI gathered from these two sources is considered as dataset~1~($DS_1$) and dataset~2~($DS_2$), respectively. 

\textbf{\textit{Gathering CTI from websites}}: We obtained a list of publicly available websites from a GitHub CTI repository~\cite{listThreatIntel2021} which are shown in Table~\ref{tab:CTIsource}. We selected these websites because they provided malware RSS feeds and their access were not restricted (e.g., API limit). We built web crawlers and scrapers to gather and extract the key pieces of information from the selected websites. A parser was built to parse the information and stored it in a structured format (i.e., a CSV file). The gathered data had consistent tagging and was labelled with the malware used in the attack, for example, Zeus, Citadel or Ice~IX. $DS_1$~contained 4060 events and represented the data available through public CTI feeds from websites. 

\textbf{\textit{Gathering CTI from MISP}}: We selected the MISP platform as a threat intelligence platform due to its popularity amongst businesses and the abundance of labelled data. We first gathered the MISP default feeds that were written in JSON format and then built a parser to extract the key attributes from that. $DS_2$~contained~213,736 events and represented the data available to an organisation from a dedicated threat intelligence platform.

\textbf{\textit{Gathering additional attributes:}} External information was gathered utilizing the parsed attributes of both~$DS_1$ and~$DS_2$. For example, the common features amongst each source were IP, domain and date. Additional attributes were gathered from WhoIS data (e.g., a database query of the RFC~3912 protocol) for each IP. We used the Python cymruwhois\footnote{https://pypi.org/project/cymruwhois/} module to search each IP in the WhoIS database, which returned the IPs ASN (i.e, a unique global identifier), owner, and country location. Besides, AlienVault forum was chosen as an external information source. We scraped the AlienVault forum updates using the Python module BeatifulSoup\footnote{https://pypi.org/project/beautifulsoup4/}. The AlienVault data, which were natural language text descriptions, was searched and extracted for the associated event and threat.

\subsection{Building Data Processor}

We built a data processor to clean, pre-process and transform the collected data for building ML-based validators. 

\textbf{\textit{Cleaning and pre-processing:}} We used the Python sci-kit learn libraries to pre-process the attribute values~\cite{chen2020deep, scikitlearn2021}. We first cleaned the data by removing null values and removing events with missing information.  We observed missing information to be relatively infrequent, resulting in minimal information loss and a more robust model. For text values, we found two types of natural language features from the MISP data (i)~text attributes which are short paragraphs that describe an event in natural language and are often taken from blogs, and (ii)~comments. 

We first analyzed the text and comments attribute to find a suitable processing and encoding technique. Thus, simple processing techniques were undertaken to decrease the dimensionality and remove any uninformative words (e.g., articles and prepositions). Each piece of text was stripped of all non-alphabetical characters, as numbers and special characters can rapidly increase dimensionality and rarely contain valuable information. The text was then stripped of any non-noun or non-proper noun tokens, as nouns are the most informative part of the text (e.g., attack names, attack types, and organisation). Finally, each word was lemmatized (i.e., changed to the base form of the word), so that similar words can be recognized. One of the key steps we followed was to tokenize the string values of attributes (i.e., domain, filename, hostname, URL) where patterns exit. We removed punctuation and special characters within a string to clean the data. We further split the text into small tokens based on a regular expression that tokenized a string using a given character. This separated each word within a value (e.g., value of domain or URL) and allowed a string to be tokenized. For example, we split the value of the URL in terms of “//” and “.”. The tokenized data were then encoded as integers to create a numeric form of a feature vector.

\textbf{\textit{Feature engineering:}} We encoded the categorical variables using one-hot encoding and label encoding. One-hot encoding considers each categorical value separately and represents each categorical variable as a column. Label encoding represents each categorical value as a unique integer. Table \ref{tab:exampleEncoding} shows an example of one-hot encoding in the first table and label encoding in the second, for three types of attacks: phishing, DDoS and SQL injection. We used the labelEncoder() method of sci-kit learn to convert the string data into numerical values. An inbuilt function from the sci-kit-learn library, standardScaler(), was used to standardize the data. The function transformed data into a normalized distribution to remove outliers from the data, allowing for building more accurate prediction models. 
The text variables (i.e., unstructured and structured natural language) did not conform to traditional one-hot or label encoding, as one-hot or label encoding interprets the text as a whole. Hence, we used two techniques: count vectorization and TFIDF as our feature engineering approach to encode text into numerical values. Count vectorization techniques stored each tokenized word as a column with its value being the number of times it appeared in each respective document.
Table~\ref{tab:exampleVectorizer} shows the examples of count vectorization of two sentences. The \textbf{TFIDF} vectorizer\footnote{https://scikit-learn.org/stable/modules/generated/sklearn.feature\_\\extraction.text.TfidfVectorizer.html} worked similar to the count vectorization, except rather than storing counts it stored the TFIDF value of each word. TFIDF provided a metric for how ‘important’ a word is within a part of the text by comparing the term’s frequency in a single document to the inverse of its frequency amongst all documents.

\begin{table}[h]
\caption{An example of one-hot encoding and label encoding for three types of attack}
    \centering
     \begin{tabular}{|c c c|c |c p{.9cm} c|}
    \hline
        Phishing & DDoS & SQL  & & & Attack & Encoded \\
                 &      & injection & & & & attack\\
        \hline
0&  1&  0&	&   1&	DDoS	& 2\\
1&	0&	0&	&	2&	Phishing&	1\\
0&	1&	0&	&	3&	DDoS&	2\\
1&	0&	0&	&	4&	Phishing&	1 \\
1&	0&	0&	&	5&	Phishing&	1 \\
0&	0&	1&	&	6&	SQL&	3\\
& & & & & injection & \\
\hline
        
    \end{tabular}
    \label{tab:exampleEncoding}
    \vspace{-10pt}
\end{table}

For the one-hot encoding setup, we used a simple count vectorizer, and for the label encoding setup we used a TFIDF vectorizer. It should be noted that whilst the dataset included text variables, the vast majority did not follow a natural language convention (e.g., domain or filename). Hence, more advanced NLP techniques, such as word embedding, cannot be accurately applied. The feature engineering schemes were saved for runtime use with the model building and prediction phase. Appendix~\ref{app:B} summarizes the pre-processing techniques that we followed for different attributes.

\begin{table*}[!hb]
\caption{Count vectors for two sentences, S1: “Fireball is malware” and S2: “Malware is any program that is harmful”}
    \centering
    \begin{tabular}{cccccccc}
    \hline
       &fireball &is &malware &any &program &that &harmful\\
       \hline
       S1 & 1 & 1 & 1 & 0 & 0 & 0 & 0\\
       \hline
       S2 & 0 & 2 & 1 & 1 & 1 & 1 & 1 \\
       \hline
   \end{tabular}
    \label{tab:exampleVectorizer}
    \vspace{-10pt}
\end{table*}

\subsection{Building the Validation Model} \label{sub:buildingValidateModel}
We built prediction models following the traditional ML pipeline (i.e., selecting ML algorithms, building prediction models, performing hyperparameter tuning and evaluating the built model). We designed a model builder to build prediction models for various attribute sets $AS$ ($ob_{attrib}$ and $un_{attrib}$). We selected eight commonly used classification algorithms~\cite{caruana2006}: Decision Tree (DT), Random Forest (RF), K-Nearest Neighbors (KNN), Support Vector Machine (SVM), Multi-Layer Perceptron (MLP), Ridge Classifier (RID), Naïve Bayes (BAY) and eXtreme Gradient Boost (XGB)~\cite{chen2016} to cover a wide range of classifier types. Appendix~\ref{app:B} summarizes the ML algorithsm that we considered to build the PoC system. Bayesian Optimization was used to automatically tune each model~\cite{snoek2012}. We used a straightforward train test split for evaluation, with~30\% of the dataset hold out for testing. In a real world, setting the training data would be selected by the threat intelligence team, to ensure data  quality. The built model was optimised by performing hyperparameter tuning. The Python module sci-kit learn  was used to build the prediction models, as it is one of the popular and widely used libraries for building prediction models~\cite{chen2020deep, scikitlearn2021, sabir2020machine}. 

\subsection{ Developing the Orchestrator}

We designed and implemented a Python script to coordinate the data collector and model builder. The script worked as an orchestrator that automated the process from gathering SOC's requirements to predicting the outputs that is validating alerts. For example, we took the SOC's requirements as an attribute set, $<ob_{atrrib}, un_{attrib}>$ and the confidence score (a value between 0 and 1). The output of the script was the value of $un_{attrib}$ and F1-score. In this process, the script first checked whether a model was available to predict $un_{attrib}$ with $ob_{attrib}$. If a model was available, it then called the data processor to process $<ob_{atrrib},\ un_{attrib}>$ and predict the value of $un_{attrib}$. 

If a model was not found, it checked the availability of CTI with attributes $<ob_{atrrib},\ un_{attrib}>$. If CTI was available, the model builder was invoked and models were built following the process of model building discussed in previous section. Here, the orchestrator used the saved feature engineering approaches and algorithm to train the model and then selected the model with the best F1-score as the optimal model. The scripts then checked the value of  the F1-score. If the F1-score was lower than the confidence score, it requested the data science team for building the model and returned that there is no model available to the security team. Otherwise, it notified the orchestrator about model availability and the next steps of data processing and predicting $un_{attrib}$ was followed and the value of $un_{attrib}$ was returned to the security team.

If required CTI data was not available, the orchestrator notified the security team about CTI unavailability. For example, the two CTI data we used did not have any vulnerability description and values. Now if we provided the vulnerability description as input and requested that we wanted to predict the severity, the orchestrator would return no data available. 

\subsection{Evaluation Metrics} \label{sub:evaluationMetrics}

Evaluation metrics are needed to measure the success of a prediction model in validating security alerts and building prediction models on run time. This will determine the effectiveness and efficiency of SmartValidator. Accuracy, precision, recall and F1-score are the four commonly accepted evaluation metrics for evaluating a prediction models performance~\cite{chen2020deep, sabir2020machine}. The correct and incorrect predictions are further calculated using number of (i)True Positive (TP) (refers to correct prediction of an attributes label), (ii) False Positive (FP) (indicates incorrect prediction of an attributes label), (iii) True Negative (TN) (refers to correct prediction that a threat does not have a particular label) and (iv) False Negative (FN) (indicates incorrect prediction that a threat does not have particular label). For example, if a model classifies a malicious IP address as non-malicious, it is calculated as a false positive. If it refers a malicious IP address as malicious it is calculated as a true positive. A true negative is when non-malicious IPs are not classified as malicious IPs. A false negative is if a malicious IP is not classified as such. Equation \ref{equ:accuracy} to equation \ref{equ:f1score} provides details of how accuracy, recall, precision and F1-score are calculated using TP, FP, TN and FN.

\vspace{-10pt}

\begin{equation}\label{equ:accuracy}
   \text{Accuracy} =  \frac{TP + TN}{TP+ TN+ FP+ FN }
\end{equation} 

\begin{equation}\label{equ:recall}
   \text{Recall}=  \frac{TP}{TP + FN}      
\end{equation}

 \begin{equation}\label{equ:precision}
     \text{Precision}= \frac{TP}{TP + FP}        
 \end{equation}  

\begin{equation} \label{equ:f1score}
    \text{F1-score}= \frac{2 \times(Recall \times Precision )}{Recall + Precision}    
\end{equation}

We assessed the effectiveness of the prediction models in validating security alerts with F1-score because  accuracy (equation \ref{equ:accuracy}) is not always a useful metric on its own. It does not capture the bias of the data. The recall (equation \ref{equ:recall}) is a measure of robustness; it displays if a model is failing to predict the relevant samples, e.g., failing to classify IP maliciousness correctly. It is important for the PoC system to have high recall to ensure that no malicious events are misinterpreted or ignored. Precision is the model's ability to accurately predict the positive class (malicious events), shown in equation \ref{equ:precision}. A low value for precision indicates a high amount of false positives. Thus, it is important to achieve high precision, as low precision would introduce the need for human validation of the output of SmartValidator. The F1-score can be considered the best metric for an overall evaluation, as it considers both precision and recall  (equation \ref{equ:f1score}) together and evaluates each class separately. F1-score does not have any unit as this is the harmonic mean of precision and recall which do not have any unit as well.

We defined a confidence score between 0-1 to be used by the security team as a threshold value for prediction models. In our PoC, we compared the confidence score with the F1-score of the prediction model. If a model had a lower F1-score than the confidence score, the PoC discarded that model.

We defined computation time, as shown in equation \ref{equ:comptime}, to evaluate the efficiency of building prediction models based on SOC's need. Computation time is the summation of the training time ($train_{time}$) and prediction time ({$predict_{time}$}). Training time is the time required to build a model and prediction time is the time required to predict unknown attributes using an optimal model.

\vspace{-15pt}

\begin{equation} \label{equ:comptime}
    computation_{time} = train_{time} + predict_{time}
\end{equation}

\vspace{-10pt}

\section {Evaluation and Results}\label{Sec:Eval}

In this section, we present the results of the developed PoC of SmartValidator to show the  effectiveness and efficiency of a dynamic ML-based validator to automate and assist the validation of security alerts with changing threat data and SOC's needs.

\begin{table*}[thb]
\caption{Performance (F1-score) of different models for prediction of \textit{attack} based on $ob^7_{attrib}$ using $DS_1$ and prediction of \textit{threat type}, \textit{threat level}, \textit{name} and \textit{event} based on $ob^{14}_{attrib}$ using $DS_2$}
    \centering
   \begin{tabular}{|l|c|c|c|c|c|}

     \hline
     
     \multirow{2}{*}{Model} & $DS_1\ ob^7_{attrib}$ & \multicolumn{4}{|c|}{$DS_2 ob^{14}_{attrib}$}  \\ [1ex]
     \cline{2-6}
     
     & attack & threat type & threat level & name & event  \\ 
     
     \hline
DT+LE & 0.719	& 0.941 & 0.998	& 0.938	& \textbf{0.995 }\\
RF+LE	& 0.274	& 0.727	& 0.76	& 0.689	& 0.362\\
KNN+LE	& 0.594	& 0.912	&	0.986	&	0.917	&	0.902\\
GBAY+LE & 0.312	& 	0.15	&	0.343	&	0.102	&	0.077\\
RID+LE	& 0.627	& 	0.055	&	0.082	&	0.105	&	0.001\\
SVM+LE	& 0.401	&	0.135	&	0.654	&	0.295	&	0.283\\
MLP+LE 	& 	0.546	&	0.762	&	-	&	0.878	&	-\\
XGB+LE	& 	\textbf{0.787}	&\textbf{	0.998}	&	\textbf{0.999}	&	\textbf{0.997}	&	-\\
\hline
DT+OHE	& 	0.587	&	0.917	&	0.988	&	0.905	&	0.926\\
RF+OHE	& 	0.501	&	0.916	&	0.948	&	0.912	&	0.87\\
KNN+OHE	& 	0.351	&	\textbf{0.998}	&	\textbf{0.998}	&	\textbf{0.999}	& \textbf{0.996}\\
GBAY+OHE	& 	0.382	&	0.677	&	-	&	0.465	&	0.241\\
RID+OHE	& 	0.763	&	0.121	&	-	&	0.007 &	- \\
SVM+OHE	& 	0.322	&	0.051	&	0.126	&	0.001	&	-\\
MLP+OHE	& 	0.762	&	0.061	&	-	&	0.079 &	-\\
XGB+OHE	& 	\textbf{0.784}	&	-	&	-	&	0.996	&	-\\
     \hline
    \end{tabular}
    \label{tab:performanceModel}
    \vspace{-10pt}
\end{table*}

\subsection{Evaluation of Effectiveness}

We evaluated the effectiveness of prediction models to answer RQ1 which is \textit{“How effective are prediction models in classifying CTI?”}. 
Specifically, we used two datasets $DS_1$ and $DS_2$, as described in Section \ref{subsec:5.2}, for predicting five observed attributes ($ob_{attrib}$). We collected the datasets in a way so that each dataset was confirmed to have at least one observed value. Finally, all experiments were conducted on the collected datasets and different combinations of attributes sets. Based on self-defined SOC requirements, CTI datasets were selected and models were built. Optimal models were selected based on their effectiveness for a particular attribute set. Effectiveness is measured using the metrics described in Section \ref{sub:evaluationMetrics}. We investigated the performance of the optimal models for classifying the threat data.

We found that 51 optimal models were returned by the PoC system based on the given requirements. Among them, seven of the models were built using $DS_1$ to predict \textit{attack} that have used $ob^1_{attrib}$ to $ob^7_{attrib}$ and the other~44 models were build using $DS_2$ to predict the four other unknown attributes based on $ob^8_{attrib}$ to $ob^{18}_{attrib}$.

The performance of different ML algorithms and encoding methods are summarized in Table~\ref{tab:performanceModel} for the two observed attributes sets – $ob^7_{attrib}$ and $ob^{14}_{attrib}$. The results shows that XGBoost (XGB) with Label Encoding (LE) achieved a near perfect F1 score while using $DS_2$. We further observed that while using One Hot Encoding (OHE), K-Nearest Neighbors (KNN) performed better than XGB. However, considering the time and memory constraints, XGB failed to train a model to predict “\textit{event}” when LE was used as an encoding method. Some of the model building processes failed as they could not finish within the allocated memory and time limits that were 24 hours and 10GB for  $DS_1$ and 48 hours and 100GB for $DS_2$. These limits were set and tested to investigate and simulate computational resource limits.  

\begin{figure*}
\centering
        \subfloat[]{\includegraphics[width=0.3\textwidth]{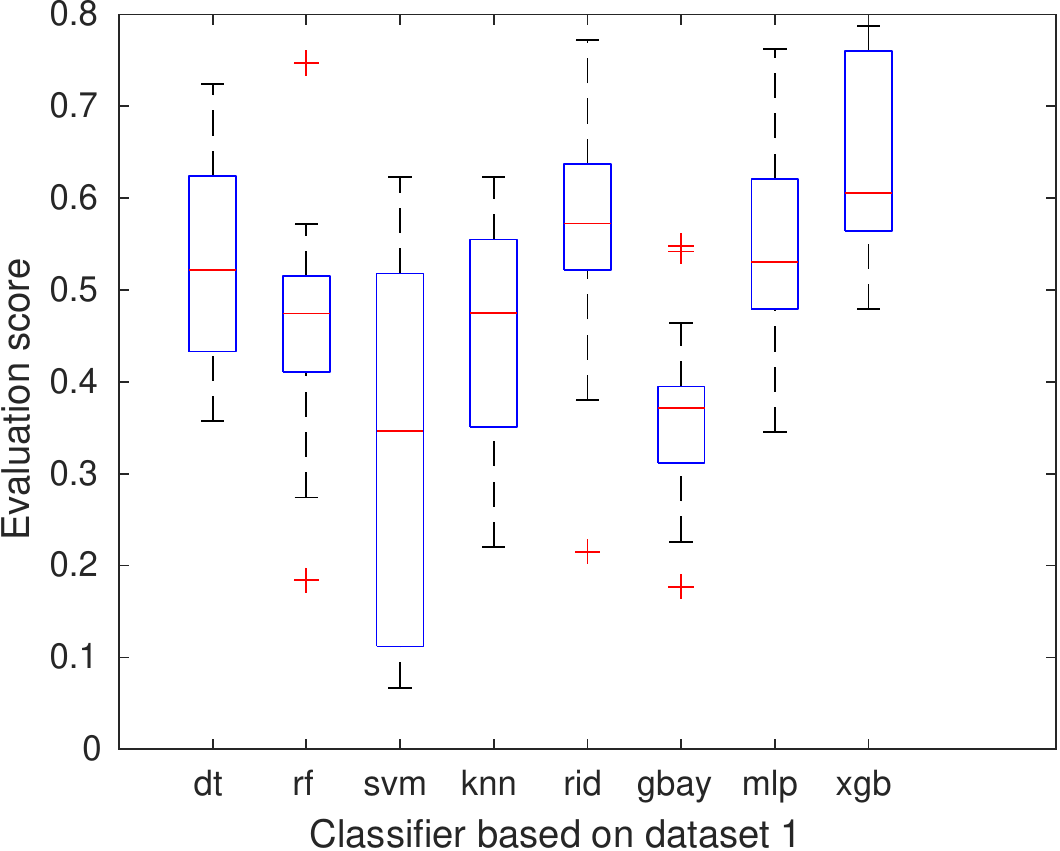}\label{fig:Effective(a)}}
        \subfloat[]{\includegraphics[width=0.3\textwidth]{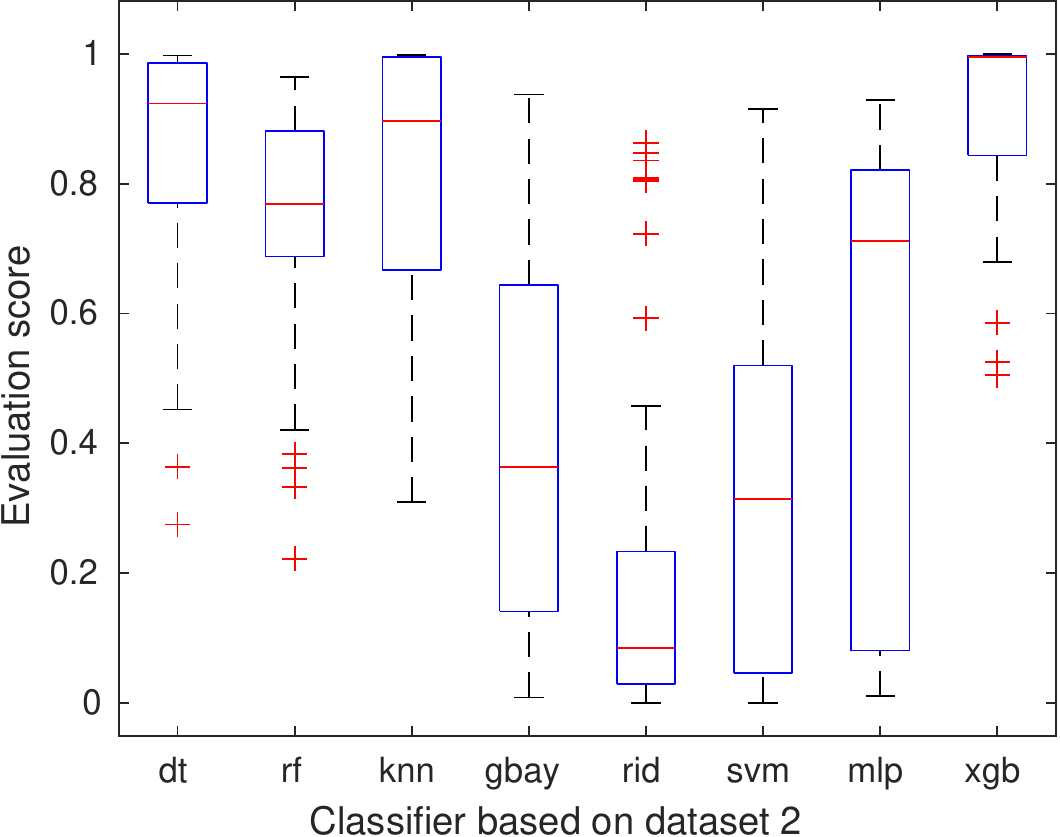}\label{fig:Effective(b)}}
        \subfloat[]{\includegraphics[width=0.3\textwidth]{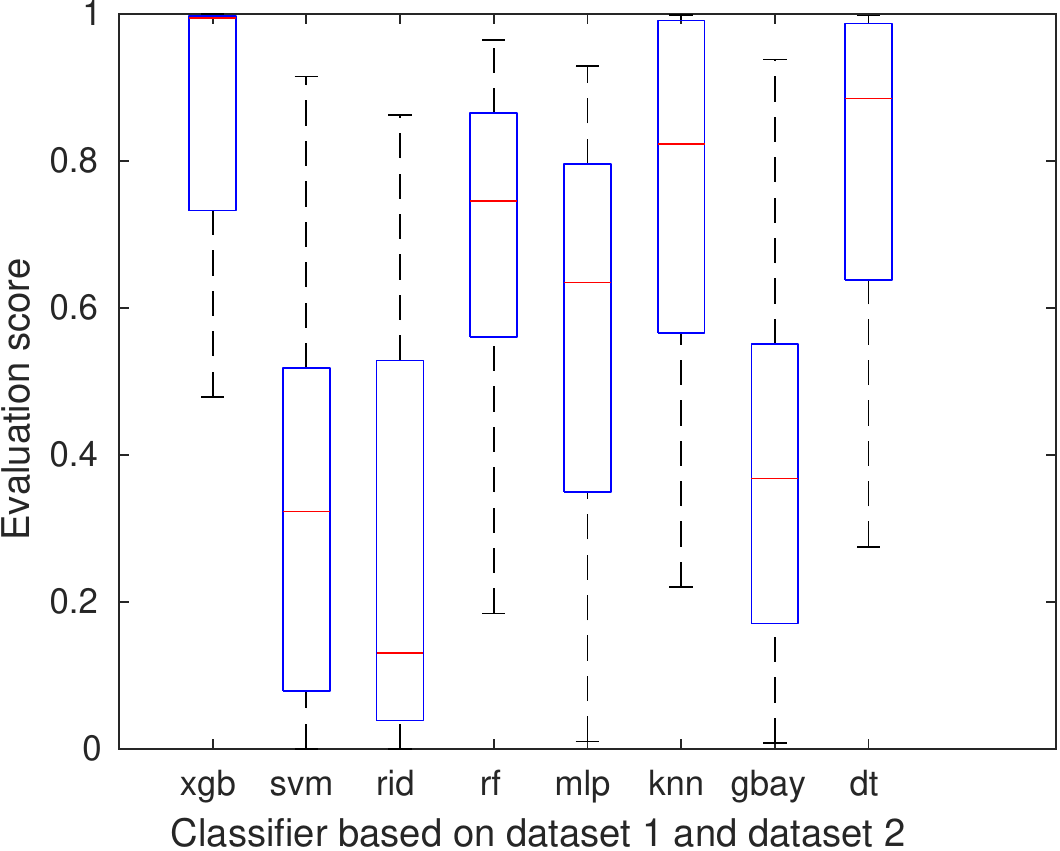}\label{fig:Effective(c)}}
        \hfil
        \subfloat[]{\includegraphics[width=0.3\textwidth]{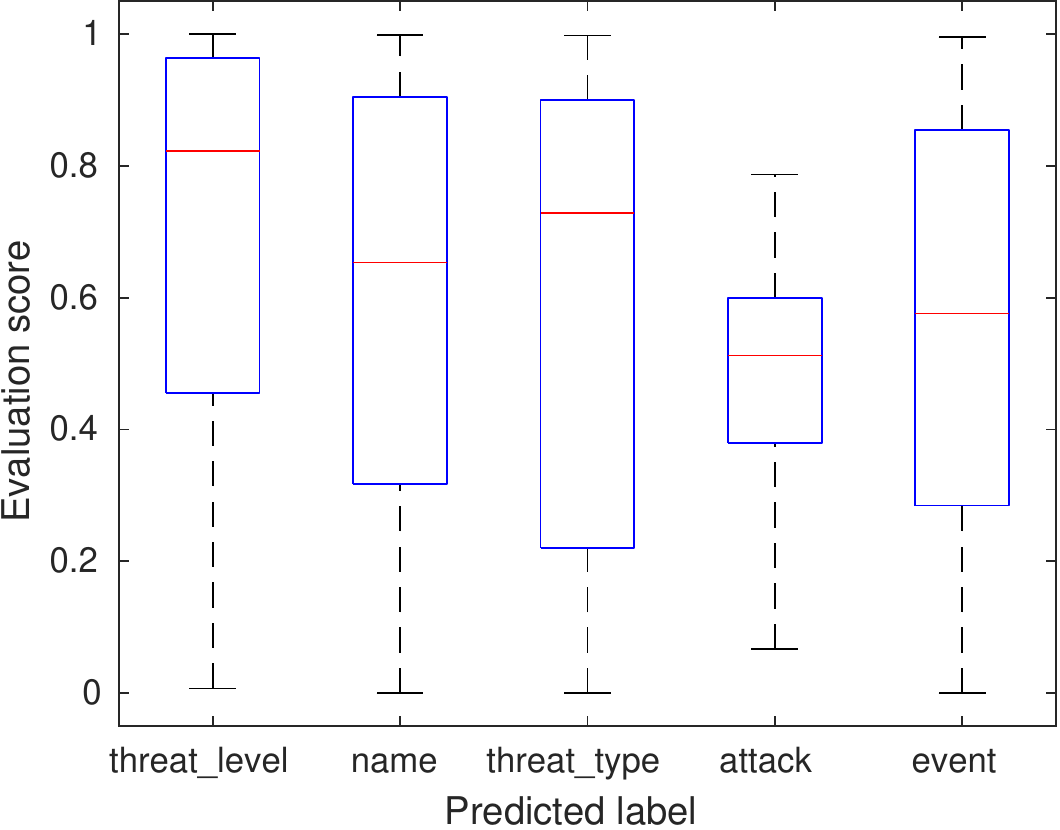}\label{fig:Effective(d)}}
        \subfloat[]{\includegraphics[width=0.3\textwidth]{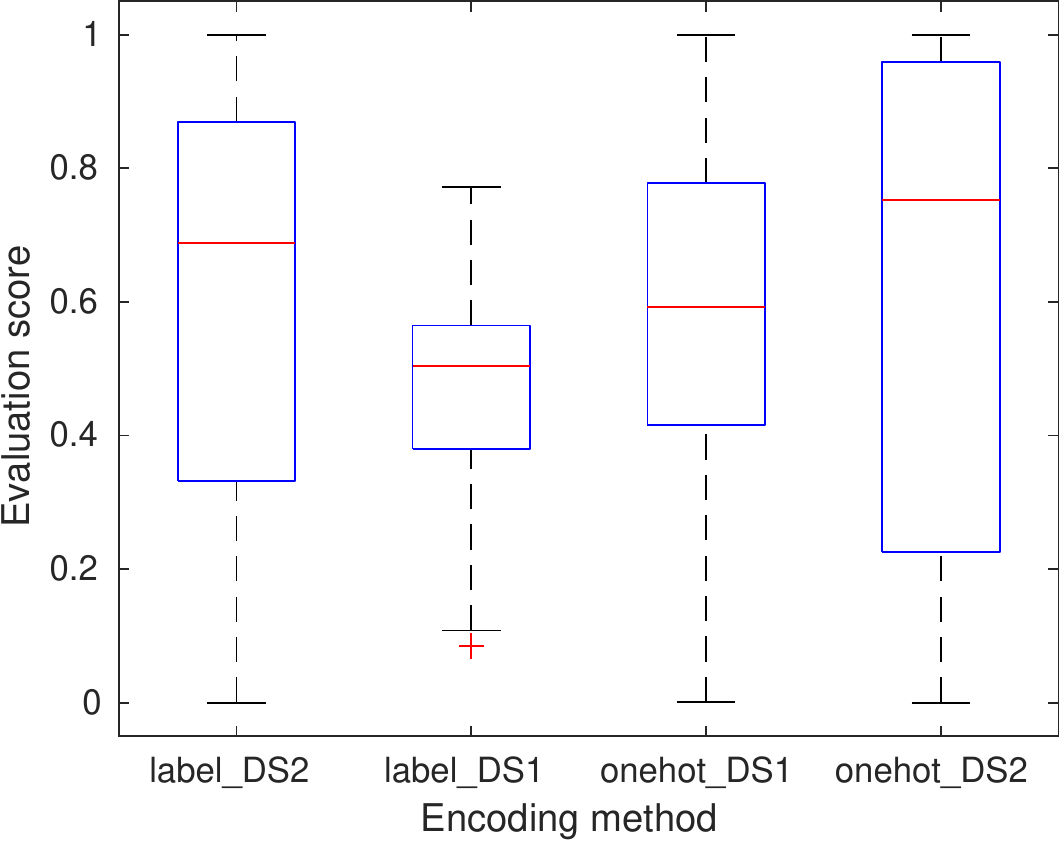}\label{fig:Effective(e)}}
        \subfloat[]{\includegraphics[width=0.3\textwidth]{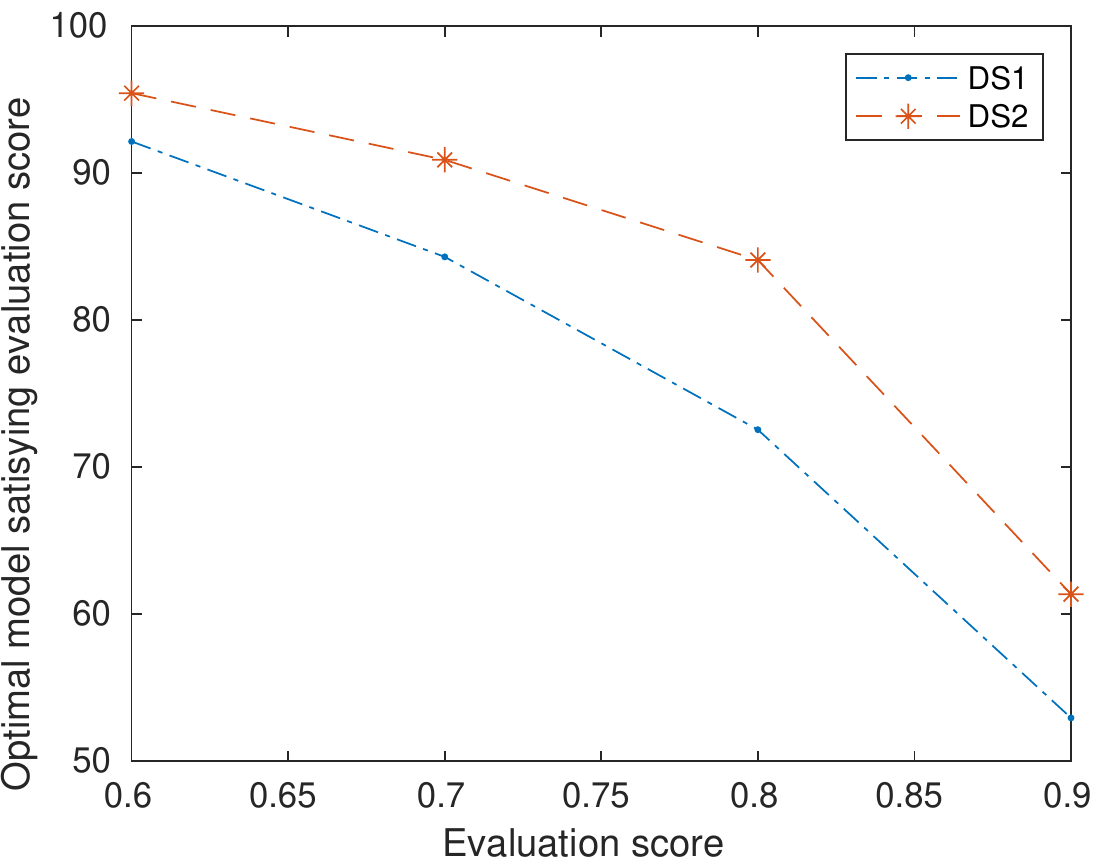}\label{fig:Effective(f)}}
        
    \caption{Comparative analysis of different classifiers performance (F1-score) while (a) using dataset 1, $DS_1$, (b) using dataset 2, $DS_2$, and (c) using both datasets, (d) predicting different labels and (e) using different encoding methods. (f) number of optimal models for different evaluation score (F1-score)}
    \label{fig:Effective}
    \vspace{-15pt}
\end{figure*}

Figure \ref{fig:Effective} shows the evaluation score (F1-score) for different datasets, labels and encoding methods. Figure \ref{fig:Effective(a)} and Figure \ref{fig:Effective(b)} shows the comparison of the different classifiers when trained on $DS_1$ and $DS_2$, respectively. We first observe that ML algorithms generally performed better using $DS_2$ data, with the exception of the Ridge classifier. This finding demonstrates the importance of CTI data and information quality. prediction models require a large number of training examples to properly learn trends and patterns. We recommend utilizing data from CTI platforms such as MISP, as these platforms aggregate a large quantity of verified information from a variety of sources.

Figure \ref{fig:Effective(c)} shows the comparative classifier performance across both $DS_1$ and $DS_2$. We observe a large range in classifier performance. The variance in best classification algorithms further motivates the need for automated model building and selection. Some ML algorithms were not as effective for this classification task. Hence, if a human were to repetitively use one algorithm to build models for different set of attributes, the results would not be good (i.e., effective) for all models. It would also be time consuming to validate every model. The results demonstrate on average, the XGB classifier performed extremely well, but KNN, as well as other tree-based classifiers (DT and RF) also performed well. MLP classifiers also appear to perform well. As MLP utilized a simplistic artificial neural network, this potentially motivates the investigation of more sophisticated deep learning methods for future work \cite{FERRAG2020102419}.

\begin{table}[]
\caption{Optimal classifier and encoding method with evaluation score (F1-score) for observed attributes $ob^1_{attrib}$ to $ob^7_{attrib}$ to predict \textit{attack} using $DS_1$}

    \centering
   \begin{tabular}{|c|c|c|}

     \hline
     
     \multirow{2}{*}{Model} & 
     \multicolumn{2}{|c|}{attack} \\
     \cline{2-3}
     
     & Optimal model & F1-score  \\ 
     
     \hline
$ob^1_{attrib}$ &	KNN + LE &	0.572 \\[1ex]
$ob^2_{attrib}$ &	RID + OHE &	0.537\\[1ex]
$ob^3_{attrib}$ &	SVM + OHE &	0.623\\[1ex]
$ob^4_{attrib}$ &	XGB + OHE &	0.728\\[1ex]
$ob^5_{attrib}$ &	RID + OHE &	0.637\\[1ex]
$ob^6_{attrib}$ &	RID + OHE &	0.772\\[1ex]
$ob^7_{attrib}$ &	XGB + LE &	0.787\\[1ex]

     \hline
    \end{tabular}
    \label{tab:optimalModelDs1}
    \vspace{-10pt}
\end{table}

We further performed comparative analysis for predicting the five observed attributes in Figure \ref{fig:Effective(d)}. Models generally performed well for all prediction tasks, but performed best when classifying \textit{threat\_level}. This is potentially because \textit{threat\_level} has the lowest dimensionality of the observed attributes, and hence ample training data for each class. Correspondingly, the \textit{event} attribute had lower performance due to its high dimensionality. $DS_1$ data was used to predict \textit{attack}. Performance is noticeably worse for this attribute,  with mean evaluation score of approximately 0.5. This further emphasizes the importance of CTI quality.

The relative effectiveness of the two encoding strategies is analyzed and shown in Figure \ref{fig:Effective(e)}. These two encoding strategies also provide a comparison of the two NLP techniques that we considered, which are count vectorization and TFIDF vectorization. One-hot encoding appeared to typically outperform label encoding, but the results are relatively similar. This refers to the similar performance of count vectorization and TFIDF, where the count vectorizer performed slightly better, as seen under the one-hot encoding results. Similar to the previous observation, we found that depending on the type of attributes and algorithms, the performance of count vectorizer and TFIDF varies. However, the performance of classification with $DS_2$, irrespective of the encoding, built effective machine learning models \cite{Helge2020, chen2020deep, islam2019multi, sabir2020machine}.  

Table \ref{tab:optimalModelDs1} shows the optimal model for predicting \textit{attack} using $ob^1_{attrib}$ to $ob^7_{attrib}$ that were built using $DS_1$. Table \ref{tab:optimalModelDs2} shows optimal model for predicting \textit{threat type}, \textit{threat level}, \textit{name} and \textit{event} based on $ob^8_{attrib}$ to $ob^{14}_{attrib}$ that were built using $DS_2$. Table \ref{tab:optimalModelDs1} and Table \ref{tab:optimalModelDs2} demonstrate that for different attribute sets different classifiers were seen to perform better. Thus, we cannot rely on a single algorithm. Table \ref{tab:optimalModelDs1} shows that the IP extracted features that were in  $ob^3_{attrib}$ (i.e., IP, ASN, owner, country) performed the best singularly, out of the three available features. $ob^5_{attrib}$ which had date and IP features was also actually relatively similar in terms of predictive capability. However, by itself the IP features that is  $ob^2_{attrib}$ can only achieve a F1-Score of 0.623. Using more available features increased the effectiveness of the prediction model, with the largest feature set achieving the best F1-score of 0.787. The addition of the domain feature to the IP feature set which formed $ob^4_{attrib}$ did not significantly increase the effectiveness of the prediction model. This is likely due to the existing correlation between the IP and domain features. However, the date did appear to noticeably improve the F1-score. The large difference between the best and worst models F1-scores highlights the importance of proper feature engineering and model selection.

Analysing the results of Table \ref{tab:optimalModelDs2}, we found that the optimal models generally performed very well, obtaining extremely good evaluation scores (F1-score). Several of the optimal models achieved a near perfect score on the testing dataset. The models also performed noticeably better than those trained on $DS_1$, further showcasing the need for good features and a large dataset. The models which used time-based features performed noticeably better than similar feature sets which did not. The time-based features likely had such strong predictive power as only a few MISP events were recorded close together. However, models without the temporal features were still able to achieve an F1-score of over 0.8. The file-based features also appeared to have weak predictive power as their inclusion or exclusion appeared to have very little impact on the evaluation score. The results of Table  \ref{tab:optimalModelDs2} further reflects that the \textit{event} label was harder to predict than other three labels. This was likely because \textit{event} was the label with the highest number of classes. Interestingly, tree-based classifiers seemed to perform better for the \textit{event} label. However, it should be noted that a lot of the more sophisticated models timed-out for the \textit{event} label, so their results were not recorded. It can also be seen that the\textit{ threat level} was the easiest to predict, likely because it had the lowest number of classes. Figure \ref{fig:Effective(d)} displays the huge variance in F1-scores of trained models for different labels.

\begin{table*}[h]
\caption{Optimal models and encoding methods with evaluation score (f1-score) for attributes sets $ob^8_{attrib}$ to $ob^{18}_{attrib}$ to predict threat type, threat level, name and event}

    \centering
   \begin{tabular}{|p{1.2cm}|l|p{.9cm}|l|p{.9cm}|l|p{.9cm}|l|p{.9cm}|}

     \hline
     
     \multirow{2}{1.2cm}{Observed attributes} & 
     \multicolumn{2}{c|}{threat type}  & \multicolumn{2}{c|}{threat level} & \multicolumn{2}{c|}{name} &\multicolumn{2}{c|}{event}\\
     \cline{2-9}
     
     & Optimal model & F1-score  & Optimal model & F1-score  & Optimal model & F1-score  & Optimal model & F1-score  \\ 
     \hline
$ob^8_{attrib}$ &	MLP + OHE&	0.648&	XGB + LE&	0.68&	RID + OHE&	0.593&	DT + LE&	0.275\\[1ex]
$ob^9_{attrib}$&	SVM + OHE&	0.854&	XGB + LE&	0.84&	SVM + OHE&	0.809&	SVM + OHE&	0.735\\[1ex]
$ob^{10}_{attrib}$	&SVM + OHE&	0.859&	SVM + OHE&	0.915&	SVM + OHE&	0.864&	SVM + OHE&	0.763\\[1ex]
$ob^{11}_{attrib}$	&KNN + OHE&	0.998&	XGB + LE&	0.999&	KNN + OHE&	0.999&	DT + LE	&0.898\\[1ex]
$ob^{12}_{attrib}$&	KNN + OHE&	0.997&	KNN + OHE&	0.999&	KNN + OHE&	0.998&	DT + LE	&0.994\\[1ex]
$ob^{13}_{attrib}$&	KNN + OHE&	0.998&	XGB + LE&	0.999&	KNN + OHE&	0.998&	KNN + OHE&	0.995\\[1ex]
$ob^{14}_{attrib}$	&KNN + OHE&	0.998&	XGB + LE&	0.999	&KNN + OHE&	0.999&	KNN + OHE&	0.999\\[1ex]
$ob^{15}_{attrib}$&	SVM + OHE&	0.873&	SVM + OHE&	0.862&	SVM + OHE&	0.813&	SVM + OHE&	0.725\\[1ex]
$ob^{16}_{attrib}$&	XGB + LE&	0.998&	XGB + LE&	1&	KNN + OHE&	0.998&	DT + LE	&0.996\\[1ex]
$ob^{17}_{attrib}$&	XGB + LE&	0.997&	XGB + LE&	0.999&	KNN + OHE&	0.999&	RF + OHE&	0.865\\[1ex]
$ob^{18}_{attrib}$&	XGB + LE&	0.998&	XGB + LE&	1&	KNN + OHE&	0.998&	DT + LE&	0.996\\[1ex]

     \hline

     \hline
    \end{tabular}
    \label{tab:optimalModelDs2}
    \vspace{-10pt}
\end{table*}

We further analyzed the classification confidence of the models of SmartValidator on the collected data. The evaluation results are visualized in Figure \ref{fig:Effective(f)}. We consider different confidence scores (0.6 – 0.9) as with the variation in attributes sets, preference of confidence score also varies. Figure \ref{fig:Effective(f)} shows that at runtime, with confidence score of 0.8, 80\% of the models that were built based on $DS_2$ fits the need of a SOC. These models were built based on the saved feature engineering and ML algorithms. Figure \ref{fig:Effective(f)} provides an overview to the security team whether they can rely on a dataset where the performance is not up to their requirements. It shows with increase in the confidence score the number of models above the confidence score decreases. Obviously, it can be seen that the models on $DS_2$ classified the attributes with a much higher confidence. We found that one of the key reasons behind this is that $DS_1$ had comparatively less data elements than $DS_2$. Thus, with $DS_2$ capturing the variation in data and correlating them provided better results than $DS_1$. The results show that approximately 84\% of the 51 optimal models had an F1-score (or confidence score) above 0.72 and 75\% of the models had F1-score above 0.8. Most of the models that were built with data gathered from CTI platforms can effectively predict $un_{attrib}$  based on $ob_{attrib}$ with a higher F1-score than the models that were built with CTI gathered from public websites. 

In summary, the MISP dataset (i.e., $DS_2$) was found to be a high-quality dataset that worked well with automated classification and thus validation of alerts. Hence, using attributes that are representative of possible threat data, prediction models can be built to effectively validate alerts with a substantial degree of accuracy, precision and recall. The results of $DS_2$ reflect that ML based validation models can be used to effectively validate the alerts with high quality CTI like $DS_2$. Model choice and alternatives are seen to be important steps to find the optimal models, as for different attributes sets the PoC has returned different models.

\vspace{-5pt}
\subsection{Evaluation of Efficiency} \label{subsec:6.2}

To demonstrate the efficiency of SmartValidator, we answer RQ2 that is “\textit{How efficient is SmartValidator in selecting and building prediction models on runtime over pre-building prediction models?}”.

In SmartValidator, we propose to build the models at run time based on SOC's requirements instead of pre-building all possible models. We considered the time to build possible model combinations of the eight aforementioned classifier algorithms as a baseline to compare the efficiency of SmartValidator. We observed that it was infeasible to pre-build models for every possible combination of features. For $DS_1$, there were 62 possible feature combinations, and for $DS_2$ there were 8190. This would increase the number of experiments for $DS_2$ to 524,160. For predicting the five unknown attributes with 18 attribute sets we would only require to run 1440 experiments; only 0.26\% of the total experiments. 
Thus, for calculating the efficiency, that is the computation time, based on SOC's requirement the PoC ran a total of 816 experiments; 112 experiments for DS1 (8 ML algorithms $\times$ 7 input attribute sets $\times$ 1 output attribute $\times$ 2 encoding methods) and 704 experiments for DS2 (8 ML algorithms $\times$ 11 input attribute sets $\times$ 4 output attributes $\times$ 2 encoding methods) to test all combinations of the 18 observed attribute sets, prediction models, encoding methods and five unknown attributes (i.e., classification labels). 

We considered the total time as a baseline to evaluate the efficiency of building the model at runtime. Here we attempted to simulate the resource limitations of model construction for a real-world environment. We considered $DS_1$ as a lightweight dataset and assigned the restrictions of 24 hours runtime and 10GB of memory, whereas $DS_2$ was a heavyweight dataset and assigned a 48 hours runtime limit and 100GB of memory. Any experiment that exceeded this run time or consumed too much memory would be aborted, as it was deemed impractical due to organisations' strict resources and fast response requirements \cite{islam2019multi, sonicwall2020}. In this result section, we reported the efficiency in terms of time.

For $DS_1$, all 112 experiments completed successfully. However, for $DS_2$, 169 of the 704 experiments failed to finish whilst enforcing our experimental setup. The most common classifier model to time out were MLP and XGB, as these models had a significantly larger training time. For these failed jobs, 149 out of 169 jobs were either for the \textit{event} or \textit{threat level} label, as these datasets had many more valid entries, and thus also took more time and memory to train. Similarly, 116 of the failed jobs used one-hot encoding, as this encoding method was much less efficient than label encoding, due to every possible value adding a dimension to the encoded input. However, 53 of the label encoding experiments also failed due to the size of the text features. These features were encoded with very high dimensionality due to the lack of a natural language convention. To investigate this issue further in our future work, we plan to investigate efficient encoding methods through vocabulary size and dimensionality reduction.  

Table \ref{tab:timeDs1} and Table \ref{tab:timeDs2} show the training time and prediction time of the optimal models that were built based on $DS_1$ and $DS_2$ respectively. The experimental results show that it is extremely inefficient to pre-build a large number of prediction models. For $DS_1$ the total training time was 61064 seconds (0.7 days), and for $DS_2$ the total training time was 7010279 seconds (81.1 days). The prediction time of a model was significantly faster than the training time, which further encouraged the use of validation models. On average, models made predictions in 0.6\% of the training time for $DS_1$ (0.09 seconds), and 2.9\% for $DS_2$ (17.29 seconds). 

\begin{table}[h]
\caption{Training time and prediction time of optimal models in \textbf{seconds} for attribute sets $ob^1_{attrib}$ to $ob^7_{attrib}$ to predict \textit{attack} using $DS_1$}

    \centering
   \begin{tabular}{|c|l|c|c|}

     \hline
     
     \multirow{2}{*}{Model} & 
     \multicolumn{3}{|c|}{attack} \\
     \cline{2-4}
     
     & Optimal model & Train time & Predict time \\ 
     
     \hline
$ob^1_{attrib}$ &	KNN + LE &	758&	0.676 \\[1ex]
$ob^2_{attrib}$ &	RID + OHE &	0.90&	0.002\\[1ex]
$ob^3_{attrib}$ &	SVM + OHE &	482	&0.001\\[1ex]
$ob^4_{attrib}$ &	XGB + OHE &	2597&	0.196\\[1ex]
$ob^5_{attrib}$ &	RID + OHE &	5.6	&0.001\\[1ex]
$ob^6_{attrib}$ &	RID + OHE &	2.2	&0.001\\[1ex]
$ob^7_{attrib}$ &	XGB + LE &	1422.9&	0.09\\[1ex]
     \hline
    \end{tabular}
    \label{tab:timeDs1}
    \vspace{-10pt}
\end{table}

\begin{table*}[h]
\caption{Training time and prediction time of optimal models in \textbf{seconds} that were built to predict threat type, threat level, name and event based on attributes $ob^8_{attrib}$ to $ob^{18}_{attrib}$ using $DS_2$ }

    \centering
  \resizebox{\textwidth}{!}{%
  \begin{tabular}{|p{1.25cm}|p{1.45cm}|p{1cm}|p{.8cm}|p{1.45cm}|p{1cm}|p{.8cm}|p{1.45cm}|p{1cm}|p{.8cm}|p{1.45cm}|p{.8cm}|p{.8cm}|}  
     \hline
     
     \multirow{2}{1.3cm}{Observed attributes} & 
     \multicolumn{3}{c|}{threat type}  & \multicolumn{3}{c|}{threat level} & \multicolumn{3}{c|}{name} &\multicolumn{3}{c|}{event}\\
     \cline{2-13}
     
     & Optimal model &	Train time  &	Predict time	& Optimal model &	train Time&	Predict time &	Optimal model &	train time &	Predict time &	Optimal model &	Train time&	Predict time \\ 
     \hline
$ob^8_{attrib}$ & 	MLP+OHE& 	166623& 	0.18& 	XGB+LE& 	111788& 	56.86& 	RID+OHE	& 10959	& 0.036& 	DT+LE& 	289.87& 	0.199\\[1ex]
$ob^9_{attrib}$& 	SVM+OHE& 	1141.1	& 0.031& 	XGB+LE& 	144473& 	139.2& 	SVM+OHE& 	925.37& 	0.016& 	SVM+OHE& 	37846	& 0.686\\[1ex]
$ob^{10}_{attrib}$& 	SVM+OHE &	1199.66 & 	0.024& 	SVM+OHE& 	804.92& 	0.006& 	SVM+OHE	& 994.43& 	0.014& 	SVM+OHE& 	27004& 	0.657\\[1ex]
$ob^{11}_{attrib}$	& KNN+OHE	& 2599.72	& 23.1& 	XGB+LE& 	19190	& 3.115& 	KNN+OHE& 	984.94& 	5.98& 	DT+LE	& 371.38& 	0.204\\[1ex]
$ob^{12}_{attrib}$& 	KNN+OHE& 	2592.79& 	22.11& 	KNN+OHE& 	16871& 	176.6& 	KNN+OHE& 	2795.24	& 14.56	& DT+LE& 	534.94& 	0.197\\[1ex]
$ob^{13}_{attrib}$	& KNN+OHE	& 3177.3& 	32.04& 	XGB+LE	& 27067& 	4.296& 	KNN+OHE& 	1178.39& 	9.311& 	KNN+OHE& 	18065& 	259.95\\[1ex]
$ob^{14}_{attrib}$	& KNN+OHE& 	2535.45& 	22.09& 	XGB+LE& 	16036& 	2.556& 	KNN+OHE& 	951.54& 	6.713& 	KNN+OHE& 	16731& 	224.98\\[1ex]
$ob^{15}_{attrib}$	& SVM+OHE	& 1081.05	& 0.015& 	SVM+OHE& 	739.58& 	0.004& 	SVM+OHE& 	1019.56& 	0.007& 	SVM+OHE& 	21275& 	0.369\\[1ex]
$ob^{16}_{attrib}$	& XGB+LE& 	10552.3& 	21.38	& XGB+LE& 	4033.701& 	1.913& 	KNN+OHE& 	899.12&	5.679&	DT+LE&	102.35& 	0.213\\[1ex]
$ob^{17}_{attrib}$ & 	XGB+LE& 	19602.7& 	19.32& 	XGB+LE& 	13950.9& 	3.938& 	KNN+OHE& 	1005.59& 	6.64& 	RF+OHE& 	7036.7& 	15.78\\[1ex]
$ob^{18}_{attrib}$	& XGB+LE& 	14257.7& 	18.57& 	XGB+LE& 	8569.79	& 3.175	& KNN+OHE	& 949.95& 	5.611& 	DT+LE	& 142.76& 	0.22\\[1ex]
     
     \hline

     \hline
    \end{tabular}
    \label{tab:timeDs2}
    \vspace{-10pt}
    }
\end{table*}

\begin{figure*}
\centering
        \subfloat[]{\includegraphics[width=0.32\textwidth]{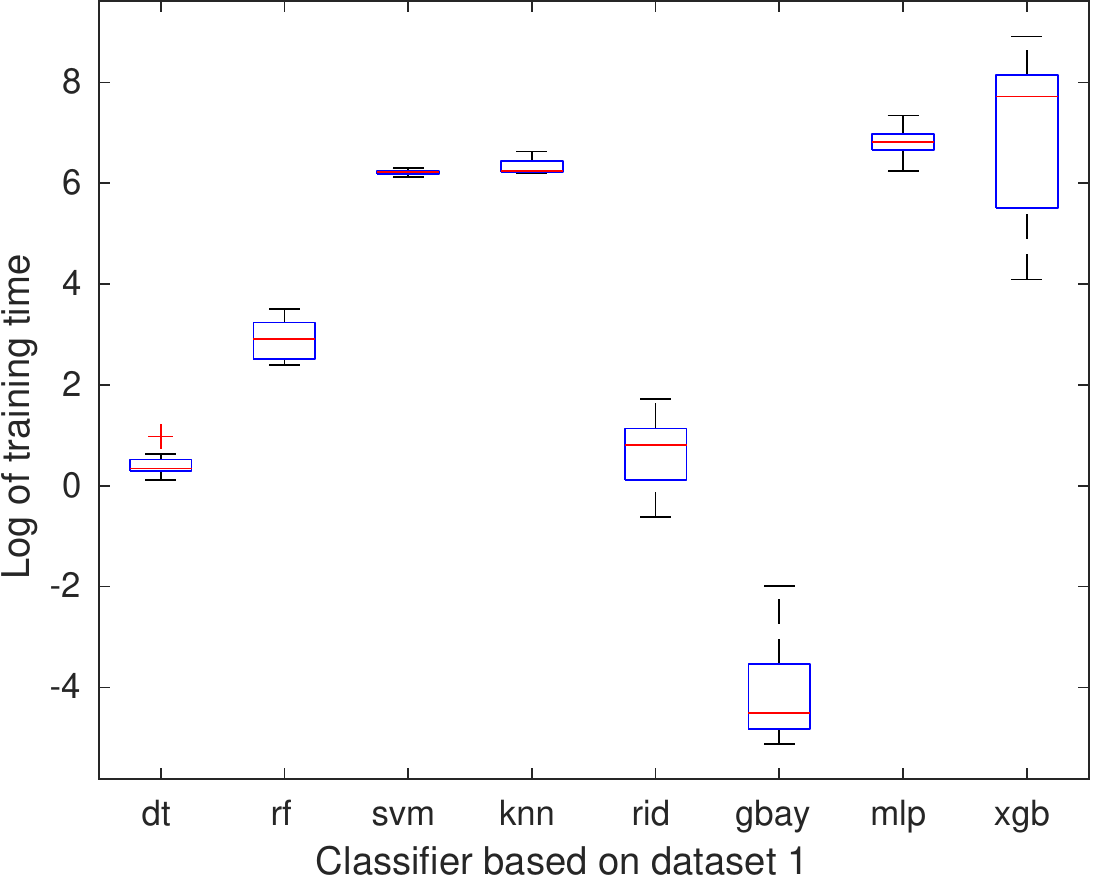}\label{fig:efficiency(a)}}
        \subfloat[]{\includegraphics[width=0.32\textwidth]{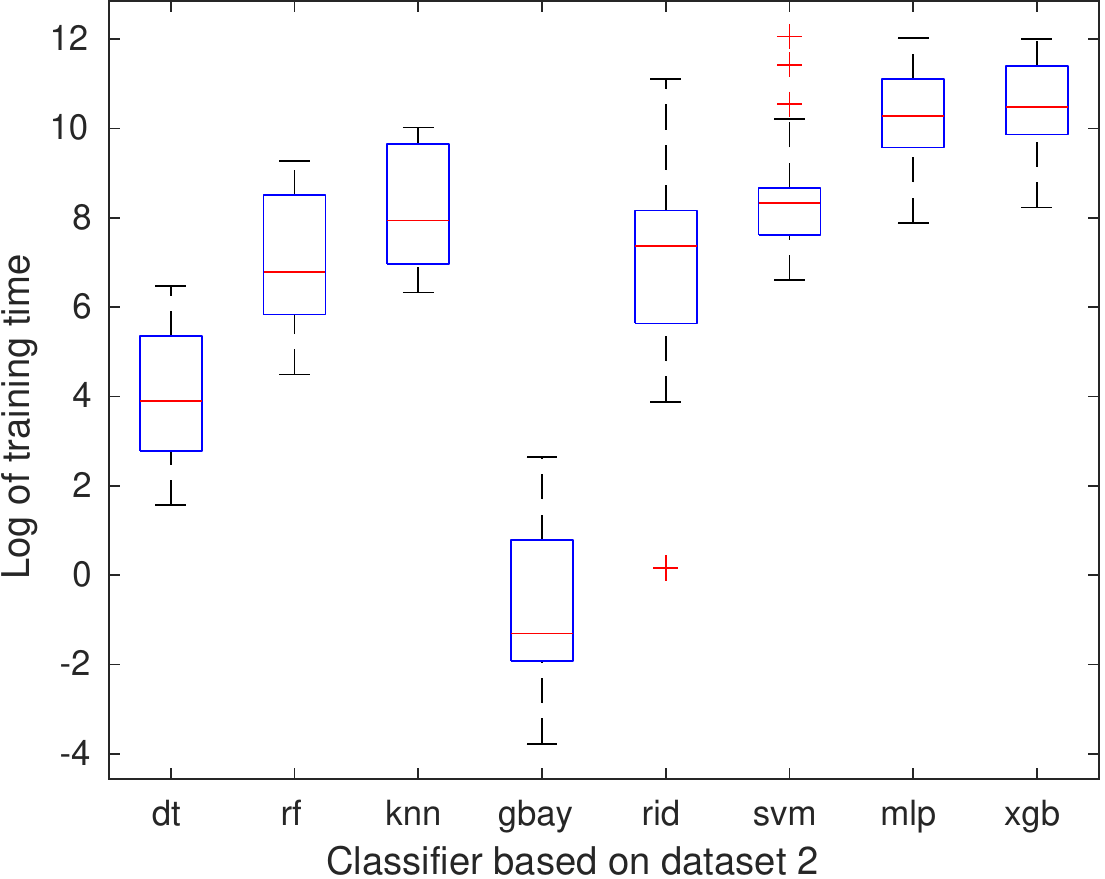}\label{fig:efficiency(b)}}
        \subfloat[]{\includegraphics[width=0.32\textwidth]{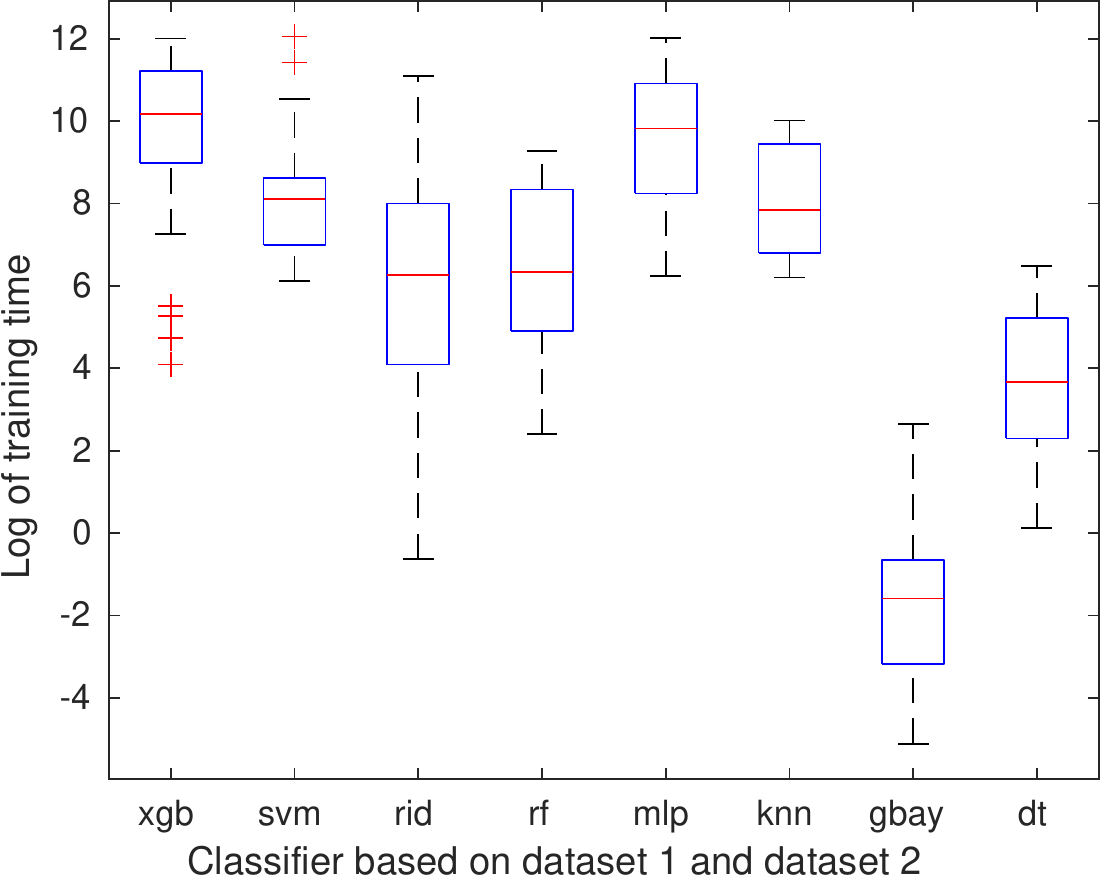}\label{fig:efficiency(c)}}
        \hfil
        \subfloat[]{\includegraphics[width=0.32\textwidth]{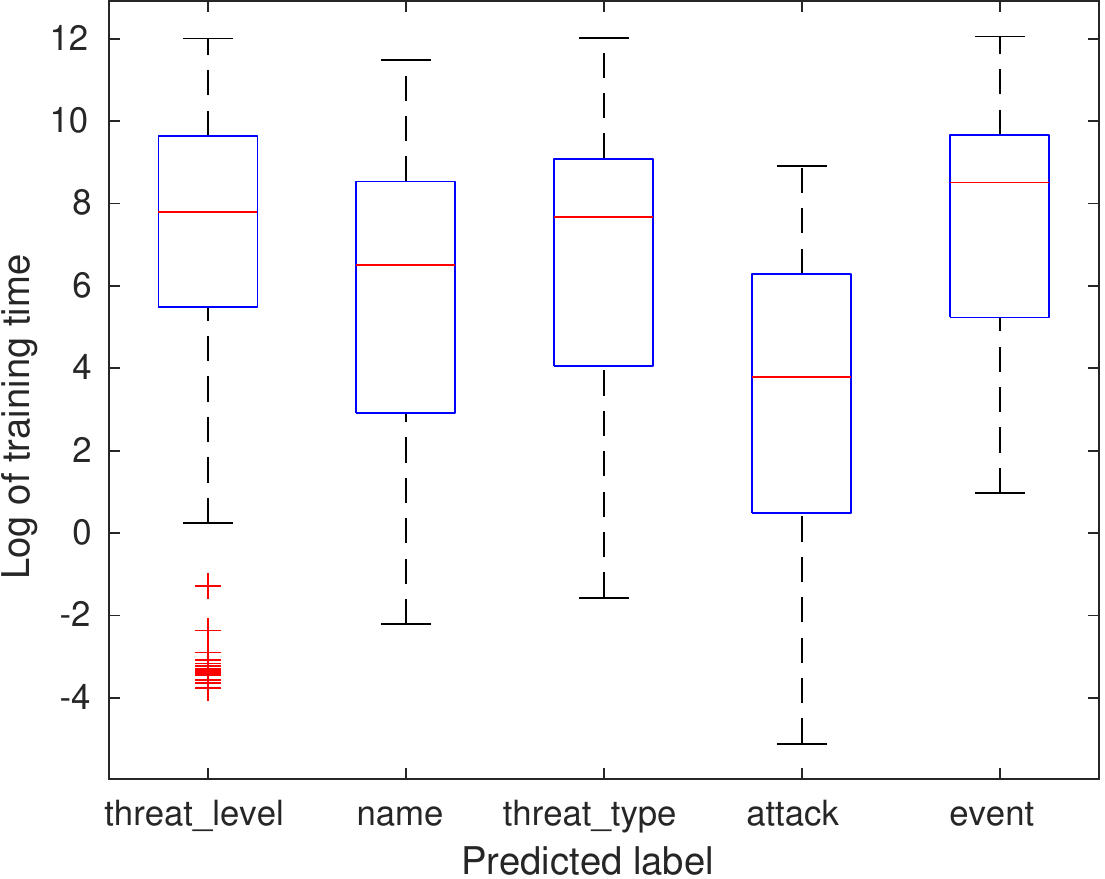}\label{fig:efficiency(d)}}
        \subfloat[]{\includegraphics[width=0.32\textwidth]{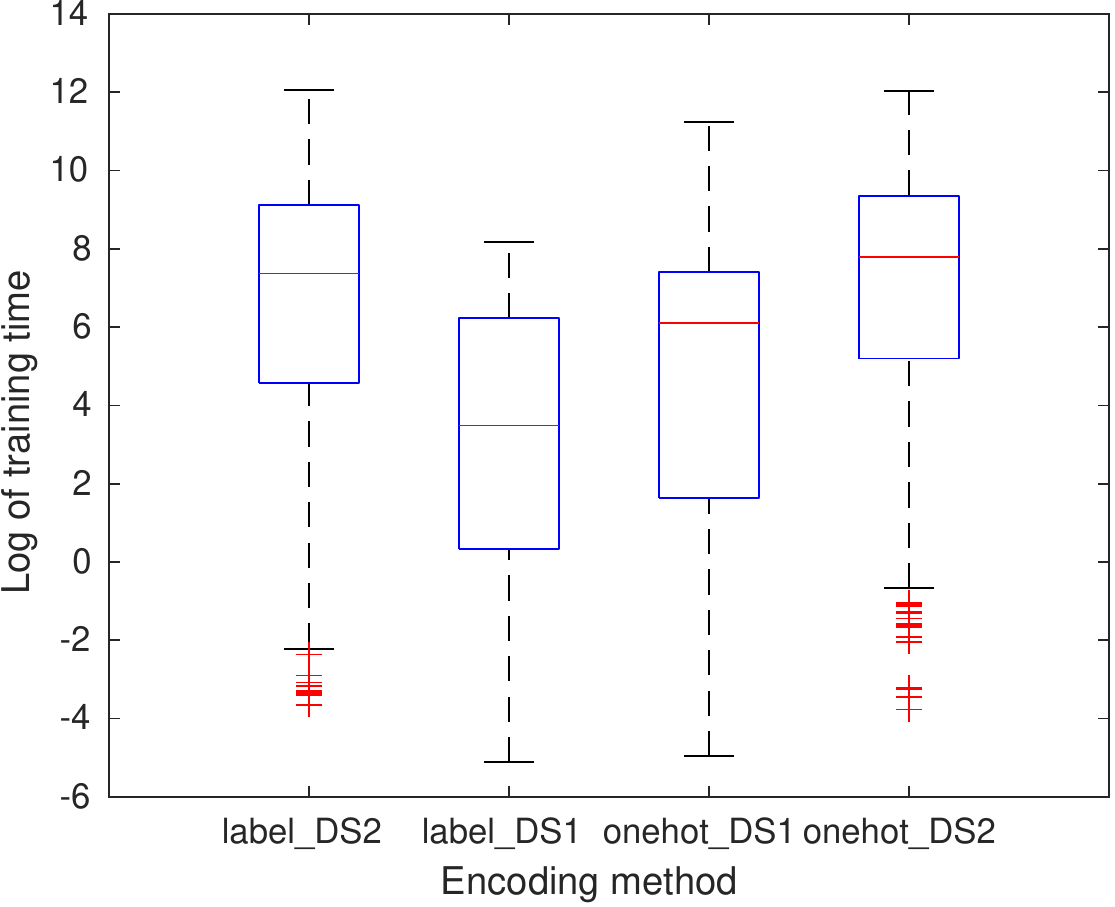}\label{fig:efficiency(e)}}
    \caption{Comparative analysis of time in \textbf{seconds} required to train different ML based validation models with (a) dataset 1, (b) dataset 2, (c) both datasets, (d) different labels and (e) encoding methods}
    \label{fig:Efficiency}
    \vspace{-15pt}
\end{figure*}

Figure \ref{fig:Efficiency} shows the logarithmic distribution of the training time for different datasets, labels and encoding methods. The logarithmic distribution of training times for $DS_1$, $DS_2$ and both $DS_1$ and $DS_2$ is shown in Figure \ref{fig:efficiency(a)}, Figure \ref{fig:efficiency(b)} and Figure \ref{fig:efficiency(c)}, respectively. It should be noted that Figure \ref{fig:Efficiency} did not consider the run time of experiments which were timed out. As shown in Figure \ref{fig:Efficiency}, the run times for more intensive models are skewed to the left. Noticeably, the Naïve Bayes (GBAY) classifiers were trained near instantaneously, as these models did not require heavy fitting to the data. DT similarly was trained faster for both datasets, due to the simplicity of this model. Figure \ref{fig:efficiency(a)} shows for $DS_1$, SVM, KNN and MLP required an average of 10-15 minutes to train. However, the XGB classifier took significantly the longest time to train with a median value of 36 minutes. For $DS_2$, the average overall training time was 217 minutes (shown in Figure \ref{fig:efficiency(b)}) which was significantly larger than the average overall training time of 9 minutes for $DS_1$, due to the substantial dataset size increase. However, XGB had a significantly larger training time of over 8 hours. We observed that the runtime between RID and SVM was quite different even though both are linear classifiers (Figure \ref{fig:efficiency(c)}). The SVM classifier took an average of 30 minutes to train on $DS_2$ due to hyperparameter optimization, whereas the RID classifier took an average of 10 seconds as they did not require any significant hyperparameters. These observations highlight the importance of SOC requirements, as we can see a trade-off between model performance and training time. XGB is the best performing model on average, but also exhibits the largest training time. Hence, SOC analysts would need to weigh model effectiveness against efficiency.

Figure \ref{fig:efficiency(d)} displays the training time for completed experiments for each predicted attribute. Models targeted towards predicting the \textit{attack} attribute only took an average training time of 9 minutes, as they were trained using the much smaller $DS_1$ dataset in comparison to the other attributes that were trained using $DS_2$. Models took an average of 2 hours to train for the name attribute, in comparison to \textit{threat\_level}, \textit{threat\_type} and \textit{event}, which took a mean time of around 4.5 hours. This could be because the training set was much smaller for the name attribute, as not as many CTI entries were assigned such information. Similarly, it should be noted that a large portion of the \textit{event} and \textit{threat\_level} experiments timed out, as the training set was larger for these attributes, due to more valid entries.

Figure \ref{fig:efficiency(e)} reflects that the training time did not significantly differ between encoding methods. This is because the major encoding dimensionality came from the domain and filename features, which were treated as text attributes and were thus only one-hot encoded in our experiments. However, one-hot encoding usually exhibited larger training times, as it has much higher dimensionality and is thus less efficient. For $DS_1$, one-hot experiments took an extra 4.6 minutes on average (11.3 minutes vs 6.8 minutes). For $DS_2$, one-hot experiments took an extra 18.9 minutes on average (227.7 minutes vs 208.8 minutes).

prediction models were built by experts (i.e., data scientists) who have the knowledge of ML technologies and pipeline to automatically validate the alerts. We observed that for changing SOC requirements, interaction or collaboration were required among the security team and the data science team where a security team specified the requirements and requested for the models they need. If the data required by the data science team were not available, they needed to request the threat intelligence team who gathered the requested information and updated the relevant list of information.
Hence, the model required redesigning and further actions needed to be performed to achieve the best prediction models.
Whilst using the PoC based on SmartVadilator, these interaction could be minimized, by managing the interaction through the orchestrator. In this way, the orchestrator requested the model builder to build the required validation model. Constructing the models automatically based on a SOC's needs required less time and was more feasible than constructing possible model for all combination of attribute sets.

Evaluating the efficiency of SmartValidator, we found that SmartValidator successfully identified and classified threat data required for alert validation. The same framework can be used to automate the validation of newly listed alerts, with new data sources. The data science team requires to map the suitable algorithm with suitable attributes sets and define the required data sources. 

\vspace{-5pt}
\subsection{Discussion}
We consider the same attributes sets and CTI sources that are used for security incident and alert validation for three validation approaches. The validation approaches are (i) manual validation, (ii) pre-building prediction models and (iii) automatic construction of prediction models based on SOC's requirements 

In manual validation, for each attribute or set of attributes, a security team first searched for attribute types and then looked for the relevant CTI availability. A security team used their previous experience to select the CTI to perform the validation. For example, to validate a malicious IP, a security team collected the blacklisted IP addresses and then looked for the IPs on the list. They further used WhoIS database to identify the relevant information about suspected IPs. The security team needed to manually write queries or call APIs to find and extract information for CTI and required knowledge about the underlying CTI sources. For similar types of alerts or changing context (i.e., change in CTI, alerts or  SOC's requirement), the same sequence of actions were repeated, that cost significant man-hours and required knowledge about the underlying plugin, API, CTI sources and so on. 

For changing context while following  approach 2 (pre-building prediction models), the security team needed to request the data science team to train and build the possible prediction models for the new context. Section \ref{subsec:6.2} reveals that to build prediction models each time a change occurs is not feasible. With automatic construction of prediction models, each time a SOC requested for a validation task, the models were built automatically. With changing context, the orchestrator coordinated the data collection and model building process that also freed the security team from coordinating and communicating with the data science team and also reduced the delay incurred due to communication. In this work, we have experimentally evaluated the performance of SmartValidator. We did not discuss the amount of time required by the security team to perform the validation activities or the time required for communication between the security team and data science team. This will also include the time gap between a request being made and the time for the security team to get the model. In future work, we plan to evaluate SmartValidator in a real SOC environment to
further demonstrate how SmartValidator can be beneficial in a SOC environment. 
For example, the effort required to manage the PoC system versus the cost saving from automation. Further we want to evaluate the maximum upfront cost to incorporate the PoC system in a real SOC environment.  

For automating construction of ML-based validation models, the PoC followed three major steps - i) collecting and processing the data, ii) training the classifier and iii) running the prediction model. Step 1 required a large amount of time  (in hours) as there were hundreds of thousands of data points to download and process, step 2 took a reasonable amount of time (in minutes) as the data were needed to be encoded and the choice of classifier were needed to train and optimise its hyperparameters. Step 3 was reasonably fast as it only needed time (in seconds) to apply a pre-trained model. These steps were bundled into an installable python package which could be made publicly available. We designed the PoC in a modular fashion so that it can be integrated into other network-enabled services to gain more information about network security. The system could easily be built for the future improvements.

To validate alerts coming from an IDS, the developed PoC system can be extended to first receive the IDS alerts over a network. After parsing the alert attributes (which would be similar to the attributes sets used in our PoC), the next steps are to transform the alert attributes into features, and then look for potential CTI to correlate the alerts information into patterns by building prediction models to predict suspicious behavior. Furthermore, the alerts can be validated using the models and the validated output can display security context of the network in a graphical user interface that is easy to understand. The PoC system can be enhanced to provide API that can be integrated as a part of a SOC's existing security system, such as middleware for EDR or SIEM. 

In our experiment, we have selected two types of CTI – one is gathered from the websites, and another is from an OSINT platform MISP, which is widely used by industry as it contains high-quality data with enriched IOCs. We consider the confidence score of users to ensure that the models with low evaluation scores are not selected. We also merged multiple data sources to enrich CTI. The experimental results show that while using MISP, SmartValidator performs better than when using web data. We assert that this is due to the high quality of MISP.  The lower level of CTIs used in our experiment can be replaced with higher quality CTIs such as TTP, where SmartValidator will perform the same steps from identifying CTIs to building models and performing the validation. The prediction models can be enriched with advanced IOC and TTP with more details about the threats.

The key steps identified to improve the models automatically built through SmartValidator are effective feature encoding, hyperparameter optimization, data distribution, feature extraction, dimensionality reduction and classifier selection. Increasing the size of the dataset and number of features increases the F1-score. The dimensionality of the categorical variables needs to be decreased. It is worth noting that our investigation is by no means exhaustive; we adopt basic ML principles and NLP techniques to develop a simplistic PoC. We plan to investigate more ML techniques such as data balancing, normalization and feature selection with diverse types of CTI. Similarly, we intend to investigate more sophisticated NLP techniques, such as word embeddings that can capture semantic information of the natural language attributes.

We found that CTI datasets contain highly multi-variate categorical variables. Highly dimensional problems like this are likely to be linearly separable, as we can separate any d+1 points in a d-dimensional space with a linear classifier, regardless of how the points are labelled. We further found that overall ensemble classifiers such as XGB performed better than the other selected seven algorithms.


As shown in Figure \ref{fig:framework}, we consider that dedicated expertise is required (e.g., a data scientist) to build prediction models, which in most cases is different from the SOC team using the models for validation task. The prediction models are built by experts (e.g., data scientists, ML experts, or developers) who are knowledgeable about ML libraries, feature engineering and algorithms. Considering a SOC's security team capability, the model building process can also be replaced with Automated Machine Learning (AutoML)\footnote{https://www.automl.org/automl/} framework such as Google Cloud AutoML\footnote{https://cloud.google.com/automl}. AutoML frameworks are designed to provide ML as a service, where a security team is required to provide the pre-processed data and for some cases the transformed data. It considers multiple ML algorithms in a pipeline to evaluate the performance, perform hyperparameter tuning and validation in an attempt to improve the performance. An AutoML framework provides a list of optimal models. For example, TPOTClassifier\footnote{http://epistasislab.github.io/tpot/api/} is an automated ML classifier that is developed in an attempt to automate the ML pipeline in python. It explores prediction models configurations that a data analyst or security analyst may not consider and attempts to fine-tune the model to achieve the most optimized model. Hence, the model builder of our proposed SmartValidator can be developed following the process discussed in Figure~\ref{fig:predictionlayer} or using an AutoML framework. Thus, depending on an organizations SOC capabilities they may use an AutoML framework instead of building the prediction models with assistance of data scientist.

SOC teams overwhelmed with massive volume of alerts failed to respond to a security incident even they had the alerts and corresponding information in their CTIs\footnote{\url{https://www.trendmicro.com/explore/en_gb_soc-research}}. Hence, we assert that the manual and repetitive validation task can be automated through SmartValidator, whereas more critical or unknown alerts and incidents would still require human involvement. In future work, we plan to extend the PoC to provide more explainable output so that SOC can make decisions based on the validated output, where prediction model choice and alternatives would be captured with explanation.

\subsection{Limitation of SmartValidator}

The experimental results show the effectiveness of SmartValidator. However, we observed several cases in which SmartValidator is unable to perform the validation.
Furthermore using CTI to validate alerts for unknown known, unknown or zero-day attacks might not always be practical. To empirically investigate this, we ran an experiment separately to explore the ability of SmartValidator to detect unknown alerts. For our experiment, we used the alert data of Snort and selected IP information that is $ob_{attrib}^3$ to predict \textit{threat level}. To validate such information, we used SmartValidator to build a model trained exclusively with the MISP data ($DS_2$), so that the Snort alert data is almost entirely unseen. The Snort alert data contained 16317 distant IPs, for which only 78 of the IPs were seen in the our MISP training dataset. The prediction model achieved an F1-score of 0.307 which implies SmartValidator has some capability for prediction of unknown alerts, albeit limited. Even though the alerts and IPs are unseen, the prediction model was still able to detect some patterns inferred from the ASN, IP owner and country attributes. We assert that due to the intelligent nature and the ability to learn the underlying semantic patterns, the models built with SmartValidator have the potential to validate some unseen values of unknown attributes (i.e., unknown attributes for which the model is built depending on the observed attributes). If the alert data is entirely unseen, i.e., the IP, ASN, owner and country are all uncontained in the training data, then SmartValidator will predict an output based on the most common value in the training data.

To validate any unknown attributes (i.e., $un_{attrib}$) with specific values, SmartValidator always needs the observed attributes as input to build a model. Therefore, even if the given value of unknown attributes is not seen in the training data, it is still possible to make a correct prediction by learning the patterns from the model building phase. For example, malicious IPs can have the same domain and threat actors. Here, an IP that has not been seen before can be identified as malicious, observing the threat actors and the domain. However, it is always possible that a trained model can fail to correctly predict IPs, which is a limitation of our proposed approach. We observered there are the cases when the observed attributes are not representative for capturing and learning patterns about the unknown attributes. SmartValidator will not be applicable to automate the validation tasks in these scenarios. Quantifiable investigation of this limitation is out of the scope of this work; but it is an exciting area of exploration for future research.

CTI is time-sensitive. Hence, SOC teams will need to update the model whenever a new CTI is available. The framework can further be extended to capture the timeliness of the CTI used for building the models and keep track of the models with up-to-date CTI. However, all the CTI will not be updated simultaneously; thus, changing all the models whenever there is an update in the CTI will not be feasible.  SmartValidator can be extended to handle this situation by retraining the model when new requests come and retraining and evaluating the available model built based on old CTI.

There can be bad quality of CTI sources which may affect the performance of SmartValidator. For example, it is possible that the models infer wrong patterns with bad quality data and consider legitimate or benign IPs as malicious. There is a need of empirical studies that focus on ensuring the quality of CTIs. However, this is not within the scope of this study.

\subsection{Threats to validity}

\textit{Construct Validity:} Our choice of data for our evaluation setup may not be suitable. We have considered CTI data to also be representative of alert attributes, as CTI is often generated from existing security alerts from external organizations. However, the classifiers have not yet been tested thoroughly with real-world internal business data. This evaluation will be attempted in future work.

\textit{Internal Validity:} A potential concern is that our models are not properly optimised. The hyperparameter tuning was performed for a specific set of configurations, as testing hyperparameters with all possible combination would take a large amount of time that may not justify here. Moreover, knowing all the combination of hyperparameters is quite im-possible. Similarly, the features that we have selected to train our models are non-exhaustive. The attribute sets selected were chosen based on the attributes used for validating alerts through security orchestration. The purpose was to show that SmartValidator can automate the construction of prediction models by identifying the CTI and the constructed prediction models can effectively validate alerts based on SOC’s requirement. The attributes list might not reflect a complete list of attributes for validating certain alerts, but our system can be easily extended to several other attack scenarios.

\textit{External Validity:} Our experiments may not generalize to other datasets. The built classifiers and prediction models were evaluated based on simulated alerts attributes sets and publicly available CTIs such as MISP. 

\vspace{-10pt}

\section{Related Work} \label{Sec:RelatedWork}

Research trends are seen in the use of Machine Learning (ML) and Deep Learning (DL) in cybersecurity domain for detection and classification of cyberattacks. Most of the existing literature focused on using AI techniques such as NLP, ML and DL tools and techniques to identify and detect cyber attacks such as malware, network intrusion, data exploitation and vulnerabilities \cite{AHMED201619, Helge2020,FERRAG2020102419, GAMAGE2020102767, GIBERT2020102526,  sabir2020machine, softVul9108283}. ML algorithms are used to extract knowledge from open source public repositories which are later used to analyze attacks, or validate alerts. Although automation has been achieved in detecting and analysis of attacks, validation of alerts and incidents still requires SOC's involvement \cite{islam2019multi}.

CTI is used by security experts of SOCs to analyze and validate alerts. To ease the use of CTI, researchers have been trying to come up with a unified structured for sharing CTI \cite{menges2019unifying, tounsi2018survey}. STIX \cite{Stixbarnum2012standardizing}, TAXII \cite{taxiiconnolly2014trusted}, CyBox \cite{barnum2012cybox} and UCO \cite{menges2019unifying} are popular among them. Use of Artificial Intelligence (AI) is encouraged for identifying, gathering and extracting CTI objects \cite{RF2019, qamar2017data, Struve2017}. Various AI tools and techniques are used for knowledge extraction, representation and analytics of CTI \cite{brazhuk2019semantic, tounsi2018survey}.  For example, Zahedi et al. \cite{zahedi2018empirical} has applied topic modelling techniques such as LDA to find the security relevant topics from open source repositories such as GitHub. Another example is a system used by EY \cite{EY2017} who mined previous threat data and then analyze it to give information on threats. Using this information, they can respond to attacks and continuously monitor a system. They place data collectors at points of high movement in a network, like a server, where a system can continuously analyze data and keep the system safe. Attacks detected can be used to harvest IOCs and analyzed to discover security issues within the network. Recorded Future (widely known CTI service providers \cite{RFID2021}) also elaborated on the fact that threat data can be found in a large variety of places such as Tweets, Facebook posts and emails. They also use AI to recognize patterns in email so that phishing emails can be found based on the information of the sender or file attached.

Recent advances in CTI  domain have drawn attention to the use of the existing knowledge to automate the manual analysis of human experts and enrichment of quality CTIs \cite{azevedo2019pure,edwards2017panning, noor2019machine, zhou2019ensemble}. For example,  vulnerability description of NVD like databases are being used to predict the severity, confidentiality and availability of threats.  Le et al. \cite{le2019automated} have used NLP and traditional ML algorithms to perform automated vulnerability assessment using vulnerability description of open source public repositories. Noor et al. \cite{noor2019machine} have used data provided by STIX and Mitre corporation to identify the documents related to attacks. Azevedo et al. have proposed a platform Pure to improve quality of CTI in the form of enriched IoCs by automatically correlating and aggregating the IoCs \cite{azevedo2019pure}. They have evaluated the performance of the proposed platform with 34 OSINT feeds gathered from MISP.
 
One recent study by Recorded Future has laid out four ways of using AI techniques to extract CTI from a detected attack \cite{Struve2017}. They have defined risk score metrics to identify malicious network activity. This extends the classification from being just about whether an attack has occurred and provides more in-depth information on the threat \cite{RFteam2018, Struve2017}. Recorded Future has used NLP to increase the range of possible data sources by removing the limit on just structured information \cite{Struve2017}. They utilize extracted text and perform classification for the language, topic and company. They have applied ML and NLP techniques to rank documents to identify malware data attacks. Their model also considers different classifications needed like scoring a risk value. They do not always use ML for scoring a risk value as they often have a rule-based system for the classifier to follow. 


Unlike the above-mentioned work, we propose SmartValidator to utilize NLP and ML techniques to assist in automating validation of security alerts and incident. To the best of our knowledge, this is the first attempt to use CTI, such as MISP data, to automate the classification and validation of security threat data based on a SOC’s preferences. Here, we have investigated how effective ML algorithms are while classifying CTI to assist in alert validation.  Unlike the existing works, where possible prediction models are pre-built, here we propose to build the models on demand. We have demonstrated the efficiency of constructing prediction models dynamically. 
\vspace{-10pt}

\section{Conclusion} \label{Sec:Conclu}

Many organizations are facing difficulty to keep pace with the changing threat landscape as security experts need to identify and analyze threat data in most circumstances. Without the indulgence of automation techniques, it is impossible to reduce the burden of analyzing the CTI to make a timely decision. 
In this work, we propose a novel framework SmartValidator, to build an effective and efficient validation tool using CTI that automates the validation of the security alerts and incidents based on SOC's preferences. Different from the manual approaches, SmartValidator is designed in a way so that SOCs can add their requirements without worrying about collecting CTI and using CTI to build a validation model. SmartValidator consists of three layers: threat data collection, threat data prediction model building and threat data validation. Different teams are responsible for updating the components of different layers, thus freeing security teams from learning data processing and model building techniques. The validation task is designed as a classification problem that leverages existing NLP and ML techniques to extract features from CTI and learn patterns for alert validation. We developed a Proof of Concept (PoC) system to automatically extract features from CTI and build prediction models based on the preferences of SOCs. A SOC's preferences are collected as a set of attributes sets:~observed and unknown attributes, where the task of the PoC is to predict unknown attributes based on observed attributes.

We have demonstrated the effectiveness of SmartValidator by predicting \textit{attack}, \textit{events}, \textit{threat type}, \textit{threat level} and \textit{name}. It collected and processed data from public websites and MISP. Next, CTI with preferred attributes sets were selected to build prediction models. Eight ML algorithms were ran to build and select the models with the highest F1-score. The best model was used to predict the unknown attributes and thus validate alerts. The developed PoC constructed validation models, and can be used to validate alerts generated by the threat detection tools and find the missing information to store the data in a structured format. The results show prediction models are effective in validating security alerts. Building prediction models at run time are more efficient then building prediction models for all possible attributes sets and~CTI. 

In future work, we plan to extend the PoC system to reduce the amount of data that is sent to the SIEM tool, thus reducing the cost of data analysis. The system can also be extended to reduce the organizational dependence on human expertise to take actions against security threats such as blocking ports or identifying maliciousness of an incident. The proposed framework can assist an organization's security team to focus on decision making, rather than manually extracting and validating security alerts and incidents. The framework can also be leveraged to provide the benefit to choose CTI suitable for an organization's application rather than using generalized CTI.
\vspace{5pt}

\textbf{Acknowledgements}

This work is supported by Cyber Security Cooperative Research Centre (CSCRC), Australia.

\vspace{-10pt}

\bibliographystyle{cas-model2-names}
\bibliography{bibfile}

\vspace{-5pt}
\appendix
\section{Appendix: Example of MISP Attribute}{\label{app:A}}
 Table \ref{tab:MISPExample} and Table \ref{tab:MISPvaluePercentage} show the key attributes of MISP and the percentage of each attribute used in this study.
\vspace{-10pt}

\begin{table*}[]
\footnotesize
    \centering
    \caption{ MISP attributes and their description used in the PoC}
    \begin{tabular}{|c|p{5cm}|p {5.4cm}|p{3.4cm}|}
    \hline
\textbf{Name} &\textbf{	Description} &	\textbf{Extraction} &	\textbf{Example} \\
\hline
\textbf{event} &	The title of the MISP event. A short natural language description of event. &	The MISP ‘Event’ feature.	& Trojanized Adobe Installer used to Install DragonOK, New Custom Backdoor \\
\hline
\textbf{threat level} &	A number between 1 to 3 assigned to indicate the threat level with 3 being the highest and 0 being undefined. &	The MISP ‘Threat Level’ feature. &	1 \\
\hline
\textbf{threat type} &	The type of threat the MISP event is about. E.g. malware, exploit-kit, tool, threat-actor. &	The ‘misp-galaxy’ tag type. E.g. ‘misp-galaxy:tool=KHRAT’. Otherwise the ‘classification’ tag value. 	& malware \\ 
\hline

\textbf{name} &	The name of the threat that the MISP event is about	& The value of the ‘misp-galaxy’ tag. &	KHRAT \\
\hline
date &	The reference date of the MISP event. &	The MISP ‘date’ feature.&	2017-03-29 \\
\hline
timestamp &	The creation time of the individual MISP attribute. Expressed in Unix time. &	The MISP ‘timestamp’ feature. &	1490818721 \\ 
\hline

ip dst &	The destination IP of the IOC. The victim’s IP. &	The MISP ‘value’ feature if the MISP ‘type’ feature equals ‘ip-dst’ or ‘ip-port’.& 	23.229.221.200 \\ 
\hline

ip src &	The source IP of the IOC. The attacker’s IP. &	The MISP ‘value’ feature if the MISP ‘type’ feature equals ‘ip-src’. Also taken from IP address lookup of domain feature (below).  &	208.91.197.46 \\ 
\hline

port&	The port on which IOC was recorded. &	The MISP ‘value’ feature if the MISP ‘type’ feature equals ‘ip-port’. &	40 \\ 
\hline

domain	& The domain of the IOC. &	The MISP ‘value’ feature if the MISP ‘type’ feature equals ‘domain’, ‘hostname’ or ‘url’. Also extracted from a domain name lookup of the ip-src feature. 	& cookie.inter-ctrip.com\\ 
\hline
file hash &	The encrypted value of a file based IOC. &	The MISP ‘value’ feature if the MISP ‘type’ feature equals ‘sha1’, ‘sha256’ or ‘md5’. 	& ffc0ebad7c1888cc4 a3f5cd86a5942014 b9e15a833e57561 4cd01a0bb6f5de2e\\ 
\hline

filename &	The filename of the IOC. &	The MISP ‘value’ feature if the MISP ‘type’ feature equals ‘filename’. &	Byebye.dll \\ 
\hline

description &	A natural language description of the MISP event. &	The MISP ‘value’ feature if the MISP ‘type’ feature equals ‘comment’ or ‘text’	& Since January of this year ...\ ...\ ... which pertained to Cambodia's country code. \\ 
\hline
comment &	A short natural language comment providing some context to the IOC. &	The MISP ‘comment’ feature.&	Sample malicious URL hosting location\\ 
\hline
    \end{tabular}
    
    \label{tab:MISPExample}
    \vspace{-10pt}
\end{table*}

\begin{table}[hbt]
    \centering
    \caption{Most common attributes of the MISP OSINT feed as of Jan. 2019}
    \begin{tabular}{ccc}
    \hline
        \textbf{ Attribute} &\textbf{Number}&	\textbf{Percentage}\\
        \hline
Hostname &	41135&	19.2\\
md5	&29085&	13.6\\
Domain&	25382&	11.9\\
sha256&	22525&	10.5\\
ip-dst&	16582&	7.8\\
sha1&	15726&	7.4\\
link&	13908&	6.5\\
url	&10036&	4.7\\
filename | sha256& 	8840&	4.1\\
ip-src	&7417&	3.5\\
file &	5324&	2.5\\
text&	905&	0.4\\
comment	& 371&	0.2\\
\hline
    \end{tabular}
    \label{tab:MISPvaluePercentage}
    \vspace{-10pt}
\end{table}

\section{Apppendix: Experiment details}\label{app:B}

\subsection{ Data pre-processing} 
 We conducted the  pre-processing using Python ‘spaCy’ module. Following are the details of pre-processing of $DS_2$.
 
 \vspace{-5pt}
 
 \begin{itemize}
\item The event’s tools/malware and threat actor are extracted from misp-galaxy event tags. Tags that are labelled with threat-actor are treated as threat actors. Otherwise, it is treated as a tool or malware.
\vspace{-7pt}
\item	The port is extracted from any IP attribute that contains a port.
\vspace{-7pt}
\item	The IP address of any hostname or domain attribute is searched using the Python socket module and included as an additional IP attribute if found. 
\vspace{-7pt}
\item	The attribute timestamp is rounded to the nearest hour to increase multiplicity.
\vspace{-7pt}
\item	The natural language attributes (text, and comment) are processed. Any non-alpha character is removed, the text is converted to lowercase, and any non-noun word is removed using the Python Spacy module.
\vspace{-7pt}
\item	The labels of attack type description are slightly cleaned for the purposes of grouping and readability: each description is stripped down to a base word so that entries describing the same attack type are grouped.
\vspace{-7pt}
\item	Non-reoccurring labels are grouped into ‘other’. 
 \end{itemize}
\vspace{-7pt}
\vspace{-5pt}

\subsection{ML Algorithms} \label{app:3.2}
Following we describes the algorithms considered as the most common and effective ML classifier.

\textbf{Decision Tree (DT):} The decision tree classifier splits the data into a series of branching nodes that end in leaves. The nodes represent a rule, e.g., a data point is red, and the leaves are a classification. Trees are fast to learn and thus applied to a very wide range of problems. 

\textbf{Random Forest (RF):} The random forest classifier uses a multitude of decision trees to create a ‘forest’. Certain trees can predict certain classes more strongly than others, so the forest picks which trees to use. 

\textbf{K-Nearest Neighbours (KNN):} K-nearest neighbours is another very simple but effective classifier. The classification of the data points is made based on the K most similar instances (or neighbours) of the data point. To do that searching is made on the entire training set to find the k most similar instances.

\textbf{Support Vector Machine (SVM)}: Support vector machine separates n features into an n-dimensional space, and attempts to identify the optimal hyperplane between them. The points surrounding the border of a feature are called support vectors which define the hyperplane. SVM is considered one of the most powerful `out-of-the-box' classifiers. 

\textbf{
Multi-Layer Perceptron (MLP)}: Multi-Layer Perceptron models use a feed-forward Artificial Neural Network (ANN). It has three layers of nodes: the input layer, a hidden layer and an output layer. Each node is considered as a neuron that uses a nonlinear activation function. Even though MLP is considered as a simplistic deep learning model, the computation of MLP is quite expensive. 

\textbf{Ridge Classifier (RID): }Ridge regression is a regression model that aims to alleviate the problems of multi-collinearity and overfitting that other regression models may have. Multi-collinearity is the existence of near-linear relationships amongst the independent variables that can distort the results. 

\textbf{Naïve Bayes (BAY}): Naïve Bayes classifiers use the Bayes theorem to predict the probability of a value to be in a class. Various distribution and statistical features are used with the classifier to build the model.
\vspace{10pt}


\printcredits

\vskip5pt

\begin{wrapfigure}{l}{1in} \includegraphics[width=1in]{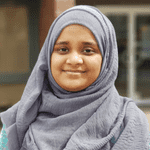} \noindent \vspace{-15pt}\end{wrapfigure} \noindent \textbf{Chadni Islam} is a postdoctoral researcher in CREST-Centre for Re-search on Engineering Software Technologies, School of Computer Science,University of Adelaide. She completed her PhD degree in 2020 under the supervision of Professor M. Ali Babar from the same school. She was co-supervised by Dr. Surya Nepal from CSIRO's Data61. Her research expertise falls at the intersection of Cyber security and Software Engineering. Her PhD research was focused on providing architecture support for security orchestration and automation systems using advanced software engineering technologies. She is interested in leveraging existing software engineering,analytical reasoning, natural language processing and machine learning tools and techniques to develop an intelligence self-adaptive security orchestration and automation platform.

\begin{wrapfigure}{l}{1in} \includegraphics[width=1in]{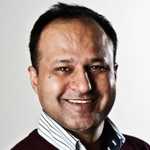} \noindent \vspace{-15pt} \end{wrapfigure}
\noindent \textbf{M. Ali Babar} is a Professor in the School of Computer Science, University of Adelaide. He is an honorary visiting professor at the Software Institute, Nanjing University, China. He has authored/co-authored more than 230 peer-reviewed papers in premier Software Technology journals and conferences. With an H-Index 48, the level of citations to his publications is among the leading Software Engineering researchers in Aus/NZ. At the University of Adelaide, Professor Babar has established an interdisciplinary research centre, CREST-Centre for Research on Engineering Software Technologies, where he leads the research and research training of more than 30 members. He has been involved in attracting several millions of dollar worth of research resources over the last ten years. Prof Babar leads the University of Adelaide’s participation in the Cyber Security Cooperative Research Centre (CSCRC), one of the largest Cyber Security initiative of the Australasian region. Within the CSCRC, he leads the theme on Platform and Architecture for Cyber Security as a Service. Further details can be found on:\url{https://researchers.adelaide.edu.au/profile/ali.babar#my-research}. 

\vspace{10pt}
\begin{wrapfigure}{l}{1in} \includegraphics[width=1in]{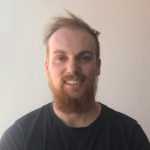} \noindent\vspace{-10pt} \end{wrapfigure} 
\noindent \textbf{Roland Croft }is currently a PhD student at the University of Adelaide, where he also completed his Bachelor's degree. He graduated with first class honours and was awarded the University Medal for achieving the highest academic score in his year. During his undergraduate study he completed several research projects in the field of cyber security. His research interests lie in natural language processing, cyber security, machine learning and data mining. His research aims to utilize open source security data and intelligence to create tools and frameworks for automated vulnerability analytic and prediction.

\begin{wrapfigure}{l}{1in} \includegraphics[width=1in]{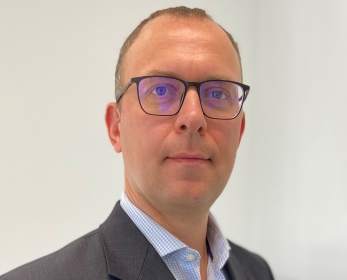} \noindent \end{wrapfigure} \noindent \textbf{Helge Janicke} is the Research Director of the Cyber Security Cooperative Research Centre, Australia. He is affiliated with Edith Cowan University and holds a visiting Professorship in Cyber Security at De Montfort University, UK. Prof. Janicke’s research interests are in the area of cyber security, in particular with applications in critical infrastructures using cyber-physical systems, SCADA and Industrial Control Systems. Prof. Janicke’s current research investigates the application of Agile Techniques to Cyber Incident Response in Critical Infrastructure, Managing Human Errors that lead to Cyber Incidents, and research on Cyber warfare \& Cyber peacekeeping. Prof. Janicke established DMU’s Cyber Technology Institute and its Airbus Centre of Excellence in SCADA Cyber Security and Forensics Research. He has been the Head of School of Computer Science at De Montfort University UK, before taking up his current position as Research Director for the Cyber Security Cooperative Research Centre. Prof. Janicke founded the International Symposium on Industrial Control System Cyber Security Research (ICS-CSR) and contributed over 150 peer reviewed articles and conference papers to the field that resulted from his collaborative research with industry partners such as Airbus, BT, Deloitte, Rolls-Royce, QinetiQ, and General-Dynamics.

\end{document}